# GLOBAL AGRICULTURAL COMPETITIVENESS INDEX (GACI) IN THE CONTEXT OF CLIMATE CHANGE: A HOLISTIC APPROACH

**Bakhtmina Zia**

Doctor of Philosophy in Economics, Institute of Management Sciences, Peshawar 25000, Pakistan

bakhtmina.zia@gmail.com

**The paper is a preprint version of the article to be published afterwards.**

**Global Agricultural Competitiveness Index (GACI) in the Context of Climate Change: A Holistic Approach**

**Abstract**

The impacts of climate change, conflicts, the spread of infectious diseases, and global economic downturns have greatly affected food production, disrupted supply chains, and hindered access to affordable, nutritious food. It poses risks to both local and global food security, in addition to agricultural market competitiveness. Given the increasing concerns about climate change and its implications for global agriculture and food security, evaluating agricultural competitiveness via a composite scale to measure the effects of climate change would be beneficial. This study examined a global agricultural competitiveness index (GACI) framework developed through a systematic review and an expert survey. The results show that most countries experienced a decline in their competitiveness scores with agricultural assessment in the context of the impact of climate change. This framework can serve as a global benchmark for assessing and comparing national and international standing. Furthermore, it can help policy development aimed at promoting sustainable and inclusive agriculture, ultimately contributing to improved global food security.

## 1. INTRODUCTION

During unstable economic times, sectors such as agriculture and food demonstrate exceptional resistance and act as equilibrating forces (Loizou et al., 2019). These sectors are vital for economic growth, employing a large portion of the population and contributing significantly to the economy (Sansika et al., 2023). According to the World Bank, agriculture is estimated to provide employment to 65% of the world's poverty-stricken adults, generate one-third of the world's gross domestic product, and feed 10 billion people by 2050 (World Bank, 2020; World Bank, 2024).

The agricultural sector faces significant risks of declining productivity, resource depletion, and environmental harm (FAO, 2021). Threats such as climate change, warfare, pests, and infectious diseases can disrupt food supply chains and hinder the availability of nutritious foods (Malik et al., 2022). Currently, there is no comprehensive measure for assessing global competitiveness in agriculture (Nugroho et al., 2023; Bobitan et al., 2023). It is imperative to address this gap by formulating a comprehensive evaluation index encompassing various dimensions of global agricultural competitiveness.

The Global Competitiveness Report indicates that a combination of institutions, policies, and factors that affect a country's productivity determine its competitiveness (Schwab & Sala-i-Martin, 2014). Embracing competitiveness can increase productivity, benefiting individuals, companies, and nations. Measuring competitiveness involves various indicators, such as productivity, cost measures, and revenue measures, and can be performed at the local, national, or regional level (Latruffe, 2010; Lei, 2023; Zia et al., 2022).

In the current world, the competitiveness of countries is linked to agricultural and food markets. Competition in these markets affects the pricing stability, accessibility, and availability of products, directly impacting farmers and food consumers. A lack of competition can hinder government initiatives aimed at these markets. The impacts on farmers may vary depending on the food security measurement tools employed for assessment (Borghi et al., 2022).

Farmers can benefit from increased competitiveness in agriculture, leading to higher returns, improved infrastructure, disaster preparedness, and increased foreign trade (Nugroho et al., 2021). Competitive markets can also enhance the quality of goods and lower consumer prices. Without a global index for measuring agricultural market competitiveness, creating one is suggested (Zia et al., 2022). Moreover, it is crucial to include climate change in this index because of its significant impact on agriculture (Tagwi, 2022; Nowak & Kasztelan, 2022).

The proposal to develop an agricultural market competitiveness index presents a prudent approach for empirical analysis. Schwab and Sala-i-Martin emphasized the importance of integrating climate change into existing competitiveness indices, recognizing it as a consequential factor (Schwab, 2011). Given the substantial impact of climate change on agriculture, failure to consider this variable in the index would result in a skewed perspective on agricultural market competitiveness. Consequently, incorporating climate change into the index would facilitate the establishment of climate-friendly policies conducive to agricultural market growth and long-term sustainability.

This paper aims to test a global agricultural competitiveness index (GACI) designed by conducting a literature review and a Delphi expert survey. The GACI has foundations in growth accounting theory, the whole-of-the-government approach and the World Economic Forum's Global Competitiveness Index (GCI). The WEF GCI lacked agricultural-specific measures for evaluating climate change impacts. The Delphi expert survey was constructed from a systematic literature review. The survey endorsed the applicability of the GCI pillars for agricultural competitiveness assessment, leading to the development of a conceptual model for GACI. The paper empirically tests the GACI framework using national data from 78 countries.

The framework will function as an international standard for evaluating and comparing agricultural status at the national and global levels and will aid in formulating policies for sustainable and inclusive agriculture, thereby enhancing global food security.

## 2. MATERIALS AND METHODS

A mixed-method study was conducted to create the Global Agricultural Competitiveness Index (GACI), which integrates twelve pillars from the World Economic Forum's (WEF) Global Competitiveness Index (GCI) with two new pillars focused on agricultural competitiveness and climate change. The accuracy of the index was verified by existing secondary data. The study utilized simple aggregation of the sub-indicators in the thirteenth pillar and performed a country-specific panel data analysis to develop the fourteenth pillar. GACI scores were calculated for 78 countries, allowing for a comparative discussion of GACI and GCI scores to assess changes in country positioning.

### 2.1. Theory and Variable Selection

The study is grounded in growth accounting theory, which analyzes the sources of economic growth by quantifying the contributions of labor, capital, and productivity. Total factor productivity (TFP) is central to this framework, representing output growth not attributable to increases in input quantities. TFP growth is linked to innovation, technological advancements, and efficiency improvements, driven by factors such as R&D, human capital, and market competition. Competitiveness refers to a country's ability to compete effectively and is influenced by productivity, innovation, infrastructure, and market conditions.

The study is based on the factors determining total factor productivity as a measure of competitiveness (Figure 1).

### 2.2 Data sources

The study employed panel data combining time series (1990--2019) and cross-sectional data from 78 countries, selected on the basis of data availability for agricultural performance and climate change impact. The GACI incorporates twelve WEF GCI pillars and two additional pillars on the basis of the literature. The 13th pillar, agricultural performance, combines agricultural total factor productivity (AGTFP), adaptation (agriculture orientation index, AOI), and the country's share in the global agricultural market, measured as agriculture, forestry and fishing value added (% of GDP) relative to world GDP.

The TFP acts as an output-to-input index for assessing agricultural productivity and changes in technical efficiency. The Agriculture Orientation Index (AOI) reflects government spending on agriculture as a percentage of GDP and offers insights into adaptation efforts. AOI can help us measure a portion of the adaptation efforts contributed by governments. Studies also provide evidence for adaptation assessment, using government actions/spending/initiatives toward the agricultural sector to promote adaptation (Luu et al., 2019). Climate change impacts are measured by the annual mean temperature and precipitation.

(Table 1).

### 2.3. Data analysis

***1. Global Agricultural Competitiveness Index:*** The GACI indicators were selected through a systematic literature review and a single-round Delphi expert survey, encompassing 12 pillars from the GCI and two novel pillars tailored for agricultural competitiveness. All 14 pillars in the GACI framework are uniformly assigned equal weights.

***2. Determinants of the GACI:*** The twelve pillars from the GCI are detailed in the **appendix (Table 1)**, while the newly designed 13th and 14th pillars are as follows;

**Pillar 13: Agriculture Performance:** Pillar 13, agricultural performance, is measured through TFP, AOI, and the country's agricultural market share, combined with equal weighting (Table 2).

***1. Total Factor Productivity:*** A key measure of agricultural competitiveness, TFP, is sourced from USDA Economic Research Service (ERS) data, with 2015 as the base year. The dataset covers the period of 1961--2019 and relies on information from the FAO, the ILO, and national agencies. The TFP is calculated as the ratio of the total output (X) to the total input (Y). Importantly, however, the USDA's TFP statistics do not account for the impact of climate change.

**Model:** TFP is the total output-to-total input ratio.

If total outputs are given by X and total inputs by Y.

Then,

$$TFP = \frac{X}{Y} \qquad (1)$$

The changes in TFP over time are expressed as follows:

$$dln\,(TFP)dt = dln\,(X)dt - dln\,(Y)dt \; (2)$$

***2. Agricultural Adaptation:*** Adaptability is necessary for competitiveness in agriculture. (Bachev, Hrabrin and Koteva, 2021) utilized the adaptability pillar to assess agricultural competitiveness, relying on microdata collected from farm managers in Bulgaria due to the lack of available data. For GACI construction, the study used the AOI to analyze adaptability in the competitiveness assessment.

**(i) Agriculture Orientation Index (AOI):** The Agriculture Orientation Index (AOI) assesses progress toward Sustainable Development Goals (SDG 2). Successful adaptation requires collaborative efforts from both the public and private sectors, which are currently underway. However, there is a lack of data on adaptation measures, particularly in agriculture, with most data being primary and in the process of being generated. While the AOI alone may not comprehensively measure adaptation efforts, it can effectively gauge the actions taken by governments in this sector. Data on nongovernmental and private sector efforts in agricultural adaptation are limited. Therefore, incorporating the AOI can assist in evaluating the portion of adaptation efforts attributed to government actions.

3. ***Country Agricultural Share in the World Market:*** In assessing a country's agricultural position in the global market, its agricultural contribution relative to the total world GDP is calculated by using value added from agriculture, forestry, and fishing. This approach allowed us to determine each country's share of agriculture in the world economy, providing insight into the economic strength and competitiveness of its agricultural sector.

Symbolically, if a country's agricultural share in the world market is shown by AgCS, then

$$AgCS = Agricultural\frac{Contribution}{World}GDP$$

where;

Agricultural contributions include agriculture, forestry, and fishing; value added; and

The world GDP is the world GDP (constant 2015 US$):

**Pillar 14: Climate Change**

***Construction of Pillar 14***

A systematic review of climate change and agricultural market competitiveness revealed that climate change is an important contributor to the turnaround of agricultural market competitiveness. Agricultural total factor productivity stands out as the most authentic measure of global agricultural competitiveness. The pillar aims to analyze the influence of climate change on overall agricultural productivity, offering insights into the competitive landscape of the national agricultural market in the face of climate-related challenges (Figure 2, Table 3).

**Model:** This study assumes a linear relationship between AgTFP and climate change. A linear regression, along with panel data analysis, is used to calculate the pillar. Country-specific effects are used to determine each country's impact.

$$AgTFP = \beta o + \beta 1temp + \beta 2prec + \beta 3countrydummy + u$$

where;;

$AgTFP$ = Annual Agricultural Total Factor Productivity

$temp$ = Annual mean temperature

$prec$ = Annual precipitation

$u$ = error term

### *3. Global Agricultural Competitiveness Index Framework*

The methodology establishes the GACI framework, which includes fourteen pillars. The twelve pillars from the GCI are recognized as relevant to agricultural markets, supported by a global Delphi expert survey that confirmed their applicability. The additional thirteenth and fourteenth pillars were developed through a comprehensive literature review and validated by the Delphi expert survey (Figure 3).

## 3. RESULTS

The Global Agricultural Competitiveness Index (GACI) was created by incorporating concepts from literature and leveraging a Delphi expert survey. It encompasses twelve pillars from the Global Competitiveness Index (GCI) of the World Economic Forum, supplemented by two new pillars tailored to assess agricultural-specific competitiveness and evaluate the influence of climate change on agriculture. The index underwent empirical testing via both secondary data and panel data analysis. GACI scores were computed for 78 countries and juxtaposed with GCI scores to gauge shifts in country rankings.

### 3.1 Pillar 13: Agriculture

In the Global Agricultural Competitiveness Index (GACI), agriculture is the 13th pillar. It is analyzed via three indicators that reflect country-specific agricultural competitiveness. These indicators are Agricultural Total Factor Productivity (AgTFP), Adaptability (Agricultural Orientation Index (AOI)), and the country's agricultural share in the world market. Each of these indicators is transformed into logarithmic form and then normalized to a range of 1--100. The arithmetic mean of these three normalized indicators is subsequently estimated to determine the overall score for this pillar. (Table 4).

### 3.2 Pillar 14 Climate Change

The fourteenth pillar of the Global Agricultural Competitiveness Index (GACI) pertains to the influence of climate change on agriculture. This assessment involves a panel regression analysis, with agricultural total factor productivity as the dependent variable and climate-related factors such as annual mean temperature, annual precipitation, and country-specific dummy variables as the independent variables**.** Robust standard errors are used to correct for heteroscedasticity (Table 5).

The analysis revealed significant impacts of climate change on agricultural total factor productivity, as evidenced by the substantial effects of temperature and precipitation. Country-specific effects were also considerable, with 76 out of 78 countries showing economically significant impacts. Specifically, 48 countries experienced negative climate impacts, whereas 30 countries experienced positive impacts. To quantify the magnitude of these impacts, constant values were added and subtracted from the country coefficients in separate columns. The resulting values were then normalized to a range of 1--100 to determine the pillar 14 scores (Table 6).

### 3.3 Global Agricultural Competitiveness Index (GACI) computed scores and rankings

The United States leads the Global Agricultural Competitiveness Index (GACI) with 81, the highest possible score. The index ranges from 32.6--81, with Mozambique having the lowest. (Table 7).

### 3.4 Global Competitiveness Index Scores and Rankings

The study utilized national data for countries with already available GCI scores. These countries were then reranked within the 78 countries (that were included in the GACI) on the basis of their GCI scores (Appendix: Table 4) and compared with the global agricultural competitiveness index (GACI) scores (Table 7). Figure 4 depicts the top ten scorers in the GCI.

### 3.5 Scorers in GACI

The top ten countries with the highest scores on the GACI are the United States, Switzerland, Sweden, Germany, the Netherlands, the United Kingdom, Denmark, Norway, France, and Austria (Figure 5).

On the other hand, the leading developing countries in the GACI are China, the Russian Federation, Chile, Poland, Malaysia, Romania, Bulgaria, Kazakhstan, Saudi Arabia, and Thailand (Figure 6).

### 3.6 Comparison between the GACI and GCI scores

The analysis of the GACI and GCI scores is shown in Figure 7. Upon computation, 74 out of the 78 countries evaluated clearly exhibit a negative disparity between the GACI and GCI scores (Appendix: Table 5). This signifies that the global agricultural competitiveness country scores are lower for most countries under review. The incorporation of climate change impacts has resulted in a reduction in countries' competitiveness scores. This underscores the fact that a failure to consider the influence of climate change on competitiveness scores may yield a distorted representation of the actual country's competitiveness standing.

An examination of the data reveals that four nations exhibit progress when the influence of climate change on agricultural competitiveness is considered. Importantly, only six developed countries experienced a decline of over 4 points in their competitiveness scores, whereas the scores for other developed countries decreased by 4 points or less. In contrast, 37 developing countries experienced a decline of over 4 points in their competitiveness scores, suggesting that climate change has a more pronounced effect on the developing world than on developed countries do.

## 4. DISCUSSION

Eight of the ten countries with the greatest difference between the GACI and GCI scores are in the Mediterranean region. Spain, Malta, Israel, Greece, and Australia are at the top of the list, with a ten-point difference in their GACI and GCI scores. Portugal, Lebanon, Albania, Morocco, and Argentina are the second-highest scorers on the list. The difference in scores is due mainly to the agroclimatic conditions and the impacts of climate change that these countries are experiencing. Unfortunately, these countries are at high risk of losing their competitiveness capability because of the severe climatic impacts they face.

The Mediterranean region is affected by ramifications of climate change, such as extended periods of drought, diminished freshwater supplies, and heightened susceptibility to desertification (Fader M, Giupponi C, Burak S, Dakhlaoui H, Koutroulis A, Lange MA, Llasat MC, Pulido-Velazquez D, 2020; Papadimitriou et al., 2016; Koutroulis et al., 2018). Research indicates that the area is prone to experiencing elevated temperatures and heat waves (Vautard et al., 2014), exacerbating soil moisture depletion (Ruosteenoja et al., 2018) and desertification hazards (Zdruli, Pandi & Cherlet, M. & Zucca, 2017). Furthermore, climate change will amplify soil erosion and wildfires, resulting in greater desertification in the region, as anticipated. Numerous studies conducted by industry leaders have focused on these critical risks (Santos et al., 2015).

The agricultural sector in the Mediterranean region will face significant challenges caused by climate change. These challenges may include lower crop productivity, higher risks of crop failure, and an increased need for irrigation (Vila-Traver et al., 2021), which could adversely affect the local economy. The critical crop development periods may experience more plant heat stress as the growing seasons become shorter. In addition, there is an increased likelihood of soil erosion and flash flooding due to more intense rainfall events during the sowing season (Zdruli, Pandi & Cherlet, M. & Zucca, 2017).

Spain, in the Mediterranean region, is notably affected by global warming. Approximately 75% of the total land area is currently at risk of desertification. The agricultural sector is particularly vulnerable to the impact of climate change, as rising temperatures can disturb the vital balance required for the growth of crops. This poses significant challenges to food production and sustainability (Sanchez, 2022).

The agri-food industry is a crucial contributor to the Spanish economy, accounting for 5.8% of the country's GDP and 11% of its overall trade. It is also among the top five exporters globally, responsible for 17% of all exports, with a trade surplus of approximately 1% of GDP and exports valued at approximately €60 billion. Climate change poses a significant threat to the agricultural industry and the economy. Global warming has resulted in several adverse effects, which have caused Spain to lose over 550 million euros or 6% of its agricultural production annually (Sanchez, 2022). As a result, Spain's GACI score has declined by ten points compared with its GCI score.

Agriculture in Malta greatly depends on irrigation, particularly for summer crops. Owing to the agroclimatic conditions of the island, there is a heavy reliance on irrigation, as precipitation is limited mainly to the semiarid season between September and March, with a value of only 10% from April to August. This results in a water shortage for crops for more than 50% of the year, with a peak in the summer (Hallett et al., 2017). Additional irrigation is necessary to ensure optimal crop growth. Although most crops in Malta are most productive during the summer, the moisture reserves of the soil have almost disappeared. This has resulted in water scarcity during crucial times for maintaining crop productivity and quality (G. et al., 2007).

The countries of Greece, Italy, Spain, and Portugal possess a notable agricultural sector that holds significant importance for both national and regional interests. Agriculture was responsible for 4% of Greece's gross value added, with some regions contributing as much as 7% to 10% in 2015 (Georgopoulou et al., 2017).

The agricultural industry plays a crucial role in feeding Australians. With over 90% of the food consumed in the country being produced domestically, agriculture, forestry, and fisheries are integral to the livelihoods of many rural communities. These sectors employ nearly 3% of all Australians and 82% of those residing in regional areas. Additionally, industry exports generate a significant portion of Australia's income, accounting for 26% of total goods and service exports in 2018--19 and 11% of all exported goods. Given that agriculture occupies more than half of Australia's land, sustainable management of arable land is crucial to ensure that industry can continue the production of the type and quality of food that Australians need (ABS, 2017; Howlett, 2021).

In recent decades, climate change has caused more severe droughts, flooding, and temperature variations to occur more frequently than in the past, adding further stress to Australian farmers (Howden et al., 2014). Agricultural practices must maximize profitability and efficiency in adapting to changing weather conditions, which are prone to market fluctuations (Freebairn, 2021).

Portugal has a highly diverse agricultural sector, with a wide range of climatic, topographical, and soil conditions. The country is classified as a climate hotspot and is one of the 26 Mediterranean areas likely to experience extreme drought worldwide. Owing to climate change, extreme weather events, such as heat waves and droughts, are becoming increasingly common. As a result, the population, economy, and agriculture are already experiencing the severe effects of drought, flooding, and wildfires, which affect many areas in the region (Schleussner et al., 2020).

Climate change will cause a decrease in crop productivity throughout southern Europe, including Portugal. Specifically, crops such as olives and grapevines, which are common in the Mediterranean region, have already experienced reductions in yield due to droughts, floods, and heat waves. This decline in food production can threaten Portugal's food security, requiring increased irrigation water to maintain crop productivity and leading to a potential water supply crisis. Additionally, agricultural operations may suffer as a result. By 2100, the value of farmland in Portugal may decrease by more than 80% (Schleussner et al., 2020; Wunderlich et al., 2023).

The impact of climate change in Lebanon cannot be overstated, particularly in rural areas where the agricultural industry is the backbone of the community. This sector is especially vulnerable to the consequences of climate change, which can have disastrous effects on crop yields, productivity, and the economy as a whole. With the country facing rising temperatures, declining rainfall, and an increasing frequency of droughts and floods, the environment and agricultural lands have already experienced significant harm. Unfortunately, farmers are struggling to adapt to the unpredictable rainy and cold seasons, which has disrupted the seasonal calendars of crops and further impacted production (Farajalla et al., 2010; Skaf et al., 2019).

Morocco is one of the most water-scarce nations in the world, and it is rapidly approaching the absolute water scarcity threshold of 500 cubic meters (m3) per person per year. Droughts are occurring more frequently and becoming more severe, contributing significantly to macroeconomic volatility and posing a risk to national food security. In the long run, climate change can cause a decline in crop yields and reduce water availability, resulting in a 6.5% GDP decline. Rain-fed agriculture, which relies on 80% of the country's land, is particularly susceptible to water shortages and droughts. Moreover, floods are a significant threat, with annual direct losses estimated to be $450 million on average. These are the most frequent climate-related natural disasters in Morocco. Another long-term stressor is sea level rise, which exacerbates the risk of flooding, especially in low-lying areas (World Bank, 2022b).

Argentina's economy heavily relies on natural resources, rendering it vulnerable to the impacts of climate change. The country boasts fertile land, propelling it to become one of the leading agricultural producers globally. However, with agro-industries contributing approximately 54% of its exports in 2021, the sector remains highly susceptible to the adverse effects of climate change (World Bank, 2022a).

The impact of climate on the agricultural industry is significant and affects both the economy and society. Droughts and excessive precipitation can reduce crop yields, impacting agricultural regions and provinces and decreasing food security. According to the (World Bank, 2021), losses in rain-fed agriculture due to water shortages or surpluses amount to an estimated $21 billion annually, or 0.61% of GDP. Agriculture accounts for approximately 60% of exports, and droughts are crucial to macroeconomic stability. In fact, more than half of the decline in economic activity in 2018 can be directly attributed to drought, which exacerbated the financial and economic crisis.

Rising temperatures and evapotranspiration would make it impossible to maintain the current 21.1 million hectares of irrigated land with existing infrastructure and water usage efficiency levels. Without intervention, climate change threatens approximately 25% of the irrigated land in the nation, resulting in $837 million in annual losses (World Bank, 2021).

Several countries, including Kazakhstan, Lesotho, Mongolia, and Russia, are improving their scores as they transition from the GCI to the GACI. Kazakhstan, in particular, is facing significant challenges in cereal production and trade due to climate change. However, increased precipitation can positively impact wheat and rice production in Kazakhstan's current water shortage situation. There are several reasons for this, one of which is that as temperatures and precipitation levels rise slightly, cereal production is likely to increase, resulting in a milder climate conducive to grain production (Yu et al., 2020).

Climatic parameters (rainfall, maximum and minimum temperatures) did not strongly affect crop production in Lesotho. However, other factors affecting crop production are those related to farmers' behavior and the types of seeds used, followed by the plowing method. Therefore, irrigated agriculture is needed to provide general stability in the food production required to match population growth (Thobei et al., 2014).

Experts predict that the effects of climate change on agriculture in Mongolia are likely detrimental, mainly because of decreased water availability, decreased soil fertility, decreased pasture productivity, and increased desertification (The World Bank Group & The Asian Development Bank, 2021). However, some studies, such as Fan's 2020 research, suggest that there may be some positive influence on agricultural production due to rising temperatures (Fan, 2020).

Russia is at the top of the countries whose competitiveness score is higher than that of the GCI. The GCI score for this country is 66.7, whereas the GACI score increases by two scores to 68.7. This finding is consistent with the findings of (Gordeev et al., 2022), who reported a significant and mostly positive impact of global climate variables on agricultural yields and harvests in Russia.

It is advisable to support financing policies and community initiatives that address the specific needs of farmers and vulnerable groups. These interventions should focus on enhancing farmers' skills and implementing technology transfer programs, providing training for diversifying livelihoods, improving extension services, and promoting the adoption of climate-smart agricultural practices to combat land degradation. It is essential to update afforestation and reforestation policies while ensuring that farmers have access to markets. Furthermore, comprehensive support for expanding irrigation schemes, enhancing farm irrigation management, and implementing smart irrigation systems is critical.

### 4.1 Region-wise country comparisons

In East Asia and the Pacific (Figure 8), certain countries are under serious threat, as evidenced by the contrast between their GCI and GACI scores. Compared with other nations, Australia has experienced the most significant decrease in its global competitiveness score, whereas China has experienced a minimal decrease. Interestingly, Mongolia's score has risen. The current scenario is particularly worrisome for Australia, given its status as a major agricultural producer and exporter. According to the Australian Bureau of Agricultural and Resource Economics and Sciences (Hughes & Gooday, 2021), agriculture accounts for 55% of Australia's land use. The decline in competitiveness may be due to the devastating impact of climate change on Australian agriculture, which has resulted in a decrease of approximately 10 points in competitiveness score.

The scores for the Eurasian countries in Figure 9 show that Portugal and Spain face the greatest decline in their competitiveness scores while moving from the GCI toward the GACI. The competitiveness score of the Russian Federation has increased by 2 points, whereas the remaining six Eurasian countries included in the study have competitiveness scores of only a single point or even less than this decline.

Among the European and North American regions (Figure 10), Malta and Albania experienced the most substantial decline in their global agricultural competitiveness index (GACI) scores, whereas Romania and Iceland experienced the lowest decrease. None of the countries have witnessed any improvements in their competitiveness scores during the GACI calculation. Notably, the competitiveness scores of nations in these regions are bearing the brunt of climate change impacts. This calls for urgent action to mitigate harmful effects and ensure a sustainable future for all.

In the Latin American and Caribbean regions (Figure 11), Argentina experienced the most significant decrease in its competitiveness score, whereas Chile's decline was comparatively minor. Nevertheless, all countries in the region experienced a decline in their competitiveness scores. Furthermore, the decline was greater throughout, with two countries experiencing a 5-point decline, four countries experiencing a 6-point decline, three countries experiencing a 7-point decline, and one country experiencing a 9-point decline in their competitiveness scores.

Within the Middle East and North Africa (Figure 12), Israel, Lebanon, and Morocco have experienced the most significant downturns in their competitiveness scores, whereas Turkey has experienced a comparatively minor decline. Bahrain and Egypt experienced a 6-point drop, Oman and Saudi Arabia experienced a 7-point decrease, and Lebanon and Morocco experienced a 9-point fall. Israel, however, has suffered the gravest challenge to its global competitiveness rankings, with a decline of 10 points.

In South Asia (Figure 13), each of the three countries examined experienced a decrease in their competitiveness scores. Nepal experienced the greatest decline, with a reduction of 7 points, followed by Pakistan, with a 6-point decrease in competitiveness score and a transition from GCI to GACI scores. Sri Lanka had the smallest decline, with a reduction of 5 points, placing them at the bottom of the list.

A study of eighteen sub-Saharan African countries, as shown in Figure 14, revealed that Botswana, Mauritius, and Rwanda experienced a maximum decline of 8 points in their competitiveness scores. Lesotho was an exception, with a slight improvement in its score. The remaining countries faced a decline in their scores, with six countries experiencing a 4-point decline, three countries experiencing a 5-point decline, one country with a 6-point decline, four countries with a 7-point decline, and three countries with an 8-point loss in their GACI scores compared with the GCI.

## 5. CONCLUSION

Competitiveness and progress are crucial factors for evaluating countries in today's global society. The agricultural and food markets are becoming increasingly competitive worldwide, posing a complex challenge when these markets are not functioning correctly. A lack of competition in these markets can lead to pricing instability, variations in price transmission, and limited accessibility and availability of products, directly affecting farmers and food consumers (FAO, 2015). Government initiatives targeting these markets may not succeed without healthy competition. Developed countries such as Australia invest significantly in promoting agricultural competitiveness. In its Delivering Ag2030 agenda, the Australian government outlines a plan to grow the agricultural sector to $100 billion by 2030 (Fell, 2022). Its strategy to improve its international agricultural competitiveness with a special focus on strengthening market access and productivity growth continues to successfully benefit its producers (Duver, A & Qin, 2020).

Enhancing agricultural competitiveness can yield significant benefits for farmers, including increased farm gate returns, improved infrastructure security and disaster preparedness, and increased foreign trade opportunities. Consumers also stand to gain from a more competitive market, as prices can be lower and quality enhanced. No comprehensive index exists to measure the competitiveness of agricultural markets, which poses a challenge for policymakers, investors, and stakeholders. A more effective and comprehensive tool is needed to provide crucial insights into the intricacies of agricultural market competitiveness and unlock opportunities for growth and development in the sector. By developing a composite index, rather than relying on individual indicators, it becomes possible to gain a more accurate understanding of the industry's competitiveness. Notably, Schwab and Sala-i-Martin (2011) underscore the need to account for climate change in such indices. Given its significant impact on agriculture, any analysis of agricultural market competitiveness that fails to incorporate climate change is inherently flawed.

Consequently, this study incorporates climate change into the index utilized to evaluate the competitiveness of agricultural markets. This approach facilitates the formulation of climate-friendly policies that are conducive to the growth and sustainability of the industry by leveraging the full potential of agricultural markets.

The Global Agricultural Competitiveness Index (GACI) is a new study that measures the competitive position of countries in the agricultural sector while accounting for climate change factors. The GACI incorporates twelve pillars from the Global Competitiveness Index (GCI) and introduces two new pillars on agriculture and climate change. On the basis of their GACI values, 78 countries were analyzed and ranked. The top performers in the GACI are the United States, Switzerland, Sweden, Germany, the Netherlands, the United Kingdom, Denmark, Norway, France, and Austria. In contrast, the leading developing countries on the GACI are China, the Russian Federation, Chile, Poland, Malaysia, Romania, Bulgaria, Kazakhstan, Saudi Arabia, and Thailand. The study reveals that only six developed countries experienced a decline of over 4 points in their competitiveness scores, whereas the scores for other developed countries decreased by 4 points or less. On the other hand, 37 developing countries faced a decline of over 4 points in their competitiveness scores. These findings indicate that agricultural vulnerability and climate change impacts are greater in the developing world than in developed countries.

## 6. POLICY IMPLICATIONS

- Global Benchmark: The GACI framework is already tested for 78 countries and has the potential to initiate a global debate, serve as a global benchmark for ranking agrarian economies, and allow countries to self-assess their respective agricultural sectors.
- Interdisciplinary Synergy: It explores the intersection of agricultural economics, environmental science, and social equity within the GACI context. This can lead to innovative theories on how social and environmental factors influence agricultural practices and competitiveness, challenging the notion that economic metrics alone can define success.
- Climate-Adapted Competitiveness Framework: It develops a theoretical model that incorporates climate resilience into the competitiveness framework. This model can define how adaptive capacities (e.g., infrastructure resilience, crop diversification) contribute to competitiveness in climate-affected regions, framing competitiveness as not only economic output but also adaptability.
- Climate-Smart Agriculture (CSA): It will promote the adoption of climate-smart agricultural practices as a means to enhance GACI rankings. For instance, offering incentives for practices like agroforestry, cover cropping, and precision farming can help countries mitigate climate impacts while boosting competitiveness.
- Policy Frameworks for Resilience: It will help recommend specific policy measures that integrate climate change adaptation into national agricultural policies. For example, governments can create subsidy programs for farmers implementing drought-resistant crops or investing in sustainable water management systems.
- GACI Climate Risk Assessment Tools: Will help develop assessment tools that incorporate climate risk factors into the existing GACI methodology. This will help countries identify vulnerabilities and opportunities for enhancing their competitiveness under changing climate scenarios.
- Collaboration Networks for Knowledge Sharing: It will pave the way into establishment of international networks for knowledge sharing among countries vulnerable to climate change. These networks will facilitate the exchange of successful adaptation strategies and technologies, enhancing overall agricultural competitiveness.
- Investment in Climate Resilience Research: It will encourage investments in research focused on climate-resilient crops and innovative farming practices, involving public-private partnerships aimed at developing new technologies that help farmers adapt to climate variability.
- Educational Initiatives: It will support the initiation and implementation of training programs for farmers on climate change impacts and adaptive strategies. For example, workshops that teach conservation tillage or integrated pest management will empower farmers to increase both resilience and competitiveness.

- Sustainability Metrics: It will lay foundations for integrating sustainability metrics into a competitiveness framework (GACI), that reflect climate change impacts. This can involve tracking carbon sequestration, biodiversity, and soil health as part of the competitiveness assessment, encouraging countries to adopt practices that enhance both productivity and environmental health.

## 7. FUTURE RESEARCH DIRECTION

- A study can be conducted on specific crops and regions to analyze the agricultural commodities in which countries/regions have specialization and a larger global market share.
- The twelve pillars of the global competitiveness index (GCI) can be redesigned to focus solely on agriculture-specific measures. However, the current study could not do so because of a lack of data and the high costs associated with data collection. Moreover, the experts surveyed in the study reported that the current pillars are equally applicable to the agricultural sector.

# TABLES

TABLE 1: Data sources

| Competitiveness Indicator | Variable | Data Source |
|---|---|---|
| Agricultural Productivity | Total Factor Productivity (TFP) | USDA |
| Adaptability (Agricultural) | Agriculture Orientation Index (AOI) | FAO STAT/WDI |
| Country Agriculture Share in World Market | Agriculture, Forestry & Fishing value added (% of GDP) | FAO STAT/WDI |
| Country Agriculture Share in World Market | World GDP | FAO STAT/WDI |
| Climate Change | Temperature | WBCCKP |
| Climate Change | Precipitation | WBCCKP |

TABLE 2: Pillar 13-Agriculture

| S.no | Indicator | Sub Indicators |
|---|---|---|
| 1 | Agricultural Productivity | Total Factor Productivity |
| 2 | Agricultural Adaptation | Agriculture Orientation Index |
| 3 | Country Agricultural Share in world market | Agriculture, forestry, and fishing, value added (constant 2015 US$)/World GDP (constant 2015 US$) |

TABLE 3: Pillar 14-Climate Change

| S.no | Indicator | Sub Indicator |
|---|---|---|
| 1 | Agricultural Productivity | Agricultural Total Factor Productivity |
| 2 | Temperature | Mean Temperature Annual |
| 3 | Precipitation | Mean Precipitation Annual |

TABLE 4: Indicators and calculation of Pillar 13-Agriculture

| Country | AgTFP | AOI | AgCS | Pillar13 |
|---|---|---|---|---|
| Albania | 30.770 | 30.097 | 34.112 | 31.660 |
| Angola | 20.559 | 32.134 | 49.440 | 34.044 |
| Argentina | 28.349 | 14.037 | 61.994 | 34.793 |
| Armenia | 61.664 | 38.899 | 28.765 | 43.110 |
| Australia | 11.946 | 53.472 | 60.933 | 42.117 |
| Austria | 42.402 | 66.382 | 41.787 | 50.190 |
| Azerbaijan | 50.716 | 64.439 | 39.622 | 51.592 |
| Bahrain | 28.771 | 73.877 | 0.751 | 34.466 |
| Botswana | 15.202 | 92.970 | 12.406 | 40.193 |
| Brazil | 40.301 | 45.772 | 72.870 | 52.981 |
| Bulgaria | 34.940 | 69.629 | 34.528 | 46.366 |
| Burkina Faso | 28.321 | 38.372 | 36.893 | 34.529 |
| Burundi | 50.650 | 25.110 | 24.770 | 33.510 |
| Chile | 37.742 | 56.062 | 49.080 | 47.628 |
| China | 38.002 | 80.808 | 100.000 | 72.937 |
| Columbia | 63.811 | 45.423 | 57.048 | 55.427 |
| Czech Republic | 28.878 | 83.877 | 41.282 | 51.346 |
| Denmark | 33.056 | 48.532 | 37.784 | 39.790 |
| Dominican Republic | 40.009 | 50.793 | 41.751 | 44.184 |
| Ecuador | 28.989 | 21.543 | 49.862 | 33.465 |
| Egypt | 40.146 | 34.362 | 65.305 | 46.604 |
| El Salvador | 42.737 | 37.304 | 28.195 | 36.078 |
| Estonia | 42.467 | 59.134 | 19.097 | 40.233 |
| France | 34.356 | 45.522 | 63.995 | 47.958 |
| Gambia, The | 0.000 | 18.263 | 11.959 | 10.074 |
| Georgia | 47.699 | 57.235 | 26.834 | 43.923 |
| Germany | 37.558 | 66.536 | 58.417 | 54.170 |
| Ghana | 38.460 | 35.676 | 51.492 | 41.876 |
| Greece | 1.211 | 34.223 | 47.489 | 27.641 |
| Guatemala | 28.686 | 39.219 | 45.621 | 37.842 |
| Guinea | 28.286 | 31.293 | 33.938 | 31.172 |
| Iceland | 100.000 | 56.931 | 25.526 | 60.819 |
| Indonesia | 48.897 | 51.694 | 77.695 | 59.429 |
| Israel | 12.792 | 58.013 | 39.506 | 36.770 |
| Italy | 22.722 | 47.592 | 63.660 | 44.658 |
| Jordan | 24.383 | 36.006 | 32.065 | 30.818 |
| Kazakhstan | 55.677 | 78.806 | 49.464 | 61.316 |
| Kenya | 18.627 | 15.428 | 53.932 | 29.329 |
| Latvia | 17.673 | 50.419 | 25.604 | 31.232 |
| Lebanon | 29.989 | 10.357 | 32.723 | 24.356 |
| Lesotho | 37.416 | 67.118 | 0.056 | 34.863 |
| Madagascar | 31.641 | 41.601 | 37.334 | 36.859 |
| Malawi | 50.290 | 66.853 | 31.841 | 49.661 |
| Malaysia | 35.066 | 47.404 | 59.993 | 47.488 |
| Mali | 32.075 | 47.056 | 44.454 | 41.195 |
| Malta | 0.826 | 81.529 | -0.000 | 27.452 |
| Mauritius | 40.110 | 72.369 | 14.755 | 42.411 |
| Mexico | 41.593 | 55.178 | 64.885 | 53.885 |
| Moldova | 59.261 | 47.722 | 26.510 | 44.498 |
| Mongolia | 53.013 | 27.774 | 31.930 | 37.573 |
| Morocco | 25.032 | -0.000 | 51.964 | 25.666 |
| Mozambique | 34.830 | 14.645 | 40.192 | 29.889 |
| Namibia | 37.816 | 51.959 | 22.421 | 37.398 |
| Nepal | 38.578 | 40.938 | 46.328 | 41.948 |
| Netherlands | 30.836 | 40.277 | 53.007 | 41.373 |
| Norway | 42.611 | 65.581 | 44.199 | 50.797 |
| Oman | 48.716 | 45.378 | 32.838 | 42.310 |
| Pakistan | 52.335 | 16.109 | 70.509 | 46.318 |
| Paraguay | 20.810 | 35.429 | 39.178 | 31.806 |
| Peru | 43.711 | 47.771 | 54.403 | 48.628 |
| Philippines | 37.116 | 50.505 | 63.379 | 50.333 |
| Poland | 38.604 | 60.908 | 50.255 | 49.922 |
| Portugal | 34.099 | 51.303 | 40.617 | 42.006 |
| Romania | 51.378 | 57.122 | 48.827 | 52.442 |
| Russian Federation | 59.323 | 57.700 | 68.529 | 61.851 |
| Rwanda | 38.424 | 44.117 | 34.668 | 39.070 |
| Saudi Arabia | 79.559 | 50.653 | 55.791 | 62.001 |
| Spain | 32.451 | 51.398 | 63.065 | 48.971 |
| Sri Lanka | 36.531 | 65.866 | 45.464 | 49.287 |
| Sweden | 57.417 | 42.754 | 46.135 | 48.769 |
| Switzerland | 33.413 | 100 | 40.608 | 58.007 |
| Thailand | 38.439 | 70.896 | 64.304 | 57.880 |
| Turkey | 42.902 | 58.248 | 69.706 | 56.952 |
| Uganda | 9.650 | 38.986 | 48.372 | 32.336 |
| Ukraine | 42.671 | 27.711 | 52.033 | 40.805 |
| United Kingdom | 30.077 | 60.951 | 56.072 | 49.033 |
| United States | 31.757 | 66.552 | 82.985 | 60.432 |
| Zambia | 44.835 | 86.838 | 25.103 | 52.258 |

TABLE 5: Climate change impact on the AgTFP-Panel Regression Results

| Agricultural TFP | Coef. | Robust St. Err. | t value | p value | [95% Conf | Interval] | Sig |
|---|---|---|---|---|---|---|---|
| Temperature | 7.935 | 1.15 | 6.90 | 0 | 5.681 | 10.189 | *** |
| Precipitation | .008 | .003 | 2.86 | .004 | .002 | .013 | *** |
| | 0 | . | . | . | . | . | |
| Albania | | | | | | | |
| Angola | -83.054 | 11.269 | -7.37 | 0 | -105.141 | -60.968 | *** |
| Argentina | -10.358 | 3.249 | -3.19 | .001 | -16.725 | -3.991 | *** |
| Armenia | 36.282 | 6.07 | 5.98 | 0 | 24.384 | 48.179 | *** |
| Australia | -67.391 | 11.406 | -5.91 | 0 | -89.746 | -45.036 | *** |
| Austria | 53.184 | 5.471 | 9.72 | 0 | 42.461 | 63.907 | *** |
| Azerbaijan | 19.521 | 1.802 | 10.83 | 0 | 15.99 | 23.053 | *** |
| Bahrain | -146.875 | 18.245 | -8.05 | 0 | -182.635 | -111.116 | *** |
| Botswana | -15.652 | 11.507 | -1.36 | .174 | -38.205 | 6.901 | |
| Brazil | -120.687 | 16.08 | -7.51 | 0 | -152.204 | -89.171 | *** |
| Bulgaria | 20.102 | 1.554 | 12.94 | 0 | 17.056 | 23.148 | *** |
| Burkina Faso | -122.202 | 19.062 | -6.41 | 0 | -159.564 | -84.841 | *** |
| Burundi | -35.091 | 10.061 | -3.49 | 0 | -54.81 | -15.372 | *** |
| Chile | 28.513 | 4.022 | 7.09 | 0 | 20.63 | 36.396 | *** |
| China | 39.895 | 5.653 | 7.06 | 0 | 28.815 | 50.976 | *** |
| Columbia | -98.298 | 16.124 | -6.10 | 0 | -129.9 | -66.695 | *** |
| Czech Republic | 40.722 | 4.141 | 9.83 | 0 | 32.605 | 48.839 | *** |
| Denmark | 38.551 | 3.999 | 9.64 | 0 | 30.713 | 46.389 | *** |
| Dominican Republic | -100.857 | 14.211 | -7.10 | 0 | -128.711 | -73.003 | *** |
| Ecuador | -82.059 | 12.108 | -6.78 | 0 | -105.789 | -58.328 | *** |
| Egypt | -69.567 | 12.382 | -5.62 | 0 | -93.835 | -45.298 | *** |
| El Salvador | -86.836 | 15.654 | -5.55 | 0 | -117.518 | -56.155 | *** |
| Estonia | 57.757 | 6.893 | 8.38 | 0 | 44.247 | 71.267 | *** |
| France | 14.995 | .747 | 20.08 | 0 | 13.531 | 16.459 | *** |
| Gambia, The | -95.399 | 18.252 | -5.23 | 0 | -131.172 | -59.626 | *** |
| Georgia | 69.913 | 5.082 | 13.76 | 0 | 59.953 | 79.872 | *** |
| Germany | 32.669 | 3.104 | 10.53 | 0 | 26.585 | 38.752 | *** |
| Ghana | -129.214 | 18.061 | -7.15 | 0 | -164.613 | -93.815 | *** |
| Greece | -8.468 | 2.559 | -3.31 | .001 | -13.484 | -3.453 | *** |
| Guatemala | -114.803 | 14.985 | -7.66 | 0 | -144.173 | -85.432 | *** |
| Guinea | -92.407 | 16.667 | -5.54 | 0 | -125.073 | -59.741 | *** |
| Iceland | 80.522 | 11.133 | 7.23 | 0 | 58.702 | 102.343 | *** |
| Indonesia | -126.339 | 17.896 | -7.06 | 0 | -161.414 | -91.264 | *** |
| Israel | -54.918 | 9.517 | -5.77 | 0 | -73.57 | -36.266 | *** |
| Italy | 4.905 | 1.031 | 4.76 | 0 | 2.884 | 6.926 | *** |
| Jordan | -52.486 | 8.336 | -6.30 | 0 | -68.825 | -36.148 | *** |
| Kazakhstan | 40.004 | 6.512 | 6.14 | 0 | 27.241 | 52.768 | *** |
| Kenya | -68.807 | 14.842 | -4.64 | 0 | -97.897 | -39.716 | *** |
| Latvia | 39.339 | 6.261 | 6.28 | 0 | 27.068 | 51.611 | *** |
| Lebanon | -2.511 | 4.441 | -0.57 | .572 | -11.214 | 6.193 | |
| Lesotho | 14.681 | .958 | 15.33 | 0 | 12.804 | 16.558 | *** |
| Madagascar | -72.217 | 12.486 | -5.78 | 0 | -96.69 | -47.745 | *** |
| Malawi | -82.106 | 12.036 | -6.82 | 0 | -105.695 | -58.517 | *** |
| Malaysia | -115.477 | 17.813 | -6.48 | 0 | -150.391 | -80.564 | *** |
| Mali | -104.631 | 19.029 | -5.50 | 0 | -141.927 | -67.335 | *** |
| Malta | -29.098 | 8.525 | -3.41 | .001 | -45.807 | -12.389 | *** |
| Mauritius | -73.467 | 13.938 | -5.27 | 0 | -100.785 | -46.149 | *** |
| Mexico | -66.103 | 10.57 | -6.25 | 0 | -86.82 | -45.386 | *** |
| Moldova | 28.621 | 2.36 | 12.13 | 0 | 23.995 | 33.247 | *** |
| Mongolia | 93.886 | 13.563 | 6.92 | 0 | 67.304 | 120.468 | *** |
| Morocco | -42.498 | 6.81 | -6.24 | 0 | -55.846 | -29.151 | *** |
| Mozambique | -61.172 | 14.112 | -4.33 | 0 | -88.831 | -33.514 | *** |
| Namibia | -42.455 | 9.624 | -4.41 | 0 | -61.317 | -23.593 | *** |
| Nepal | -8.864 | 1.79 | -4.95 | 0 | -12.373 | -5.356 | *** |
| Netherlands | 30.211 | 2.139 | 14.13 | 0 | 26.019 | 34.402 | *** |
| Norway | 59.262 | 11.35 | 5.22 | 0 | 37.016 | 81.507 | *** |
| Oman | -83.642 | 15.777 | -5.30 | 0 | -114.566 | -52.719 | *** |
| Pakistan | -54.795 | 10.068 | -5.44 | 0 | -74.527 | -35.063 | *** |
| Paraguay | -92.331 | 13.504 | -6.84 | 0 | -118.799 | -65.863 | *** |
| Peru | -59.173 | 9.28 | -6.38 | 0 | -77.362 | -40.984 | *** |
| Philippines | -112.126 | 17.088 | -6.56 | 0 | -145.618 | -78.634 | *** |
| Poland | 39.087 | 4.093 | 9.55 | 0 | 31.065 | 47.11 | *** |
| Portugal | -30.5 | 4.081 | -7.47 | 0 | -38.499 | -22.501 | *** |
| Romania | 37.092 | 2.907 | 12.76 | 0 | 31.395 | 42.789 | *** |
| Russian Federation | 137.872 | 18.869 | 7.31 | 0 | 100.888 | 174.855 | *** |
| Rwanda | -14.241 | 8.569 | -1.66 | .097 | -31.037 | 2.554 | * |
| Saudi Arabia | -72.977 | 15.154 | -4.82 | 0 | -102.678 | -43.276 | *** |
| Spain | -7.513 | 2.315 | -3.25 | .001 | -12.051 | -2.976 | *** |
| Sri Lanka | -121.383 | 18.022 | -6.74 | 0 | -156.705 | -86.061 | *** |
| Sweden | 88.149 | 10.664 | 8.27 | 0 | 67.249 | 109.05 | *** |
| Switzerland | 56.394 | 6.755 | 8.35 | 0 | 43.154 | 69.634 | *** |
| Thailand | -115.024 | 17.329 | -6.64 | 0 | -148.988 | -81.06 | *** |
| Turkey | 11.822 | 1.424 | 8.30 | 0 | 9.03 | 14.614 | *** |
| Uganda | -52.025 | 13.145 | -3.96 | 0 | -77.789 | -26.262 | *** |
| Ukraine | 18.744 | 3.706 | 5.06 | 0 | 11.48 | 26.009 | *** |
| United Kingdom | 37.663 | 3.267 | 11.53 | 0 | 31.261 | 44.066 | *** |
| United States | 38.354 | 3.247 | 11.81 | 0 | 31.99 | 44.717 | *** |
| Zambia | -76.079 | 11.748 | -6.48 | 0 | -99.104 | -53.053 | *** |
| Constant | -22.809 | 14.668 | -1.56 | .12 | -51.558 | 5.94 | |

| | | | |
|---|---|---|---|
| Mean dependent var | 91.258 | SD dependent var | 20.109 |
| Overall r-squared | 0.531 | Number of obs | 2340 |
| Chi-square | . | Prob > chi2 | . |
| R-squared within | 0.080 | R-squared between | 1.000 |

**** p<.01, ** p<.05, * p<.1*

TABLE 6: Pillar 14 country score calculation and normalization.

| Country | Coefficient | Constant | D1 = coef – cons | D2 = coef + cons | D1 Normalized | D2 Normalized |
|---|---|---|---|---|---|---|
| Albania | . | -22.809 | -22.809 | 22.809 | 51.581 | 51.581 |
| Angola | -83.054 | -22.809 | -105.864 | -60.245 | 22.413 | 22.413 |
| Argentina | -10.358 | -22.809 | -33.167 | 12.451 | 47.943 | 47.943 |
| Armenia | 36.281 | -22.809 | 13.472 | 59.091 | 64.323 | 64.323 |
| Australia | -67.391 | -22.809 | -90.200 | -44.582 | 27.914 | 27.914 |
| Austria | 53.184 | -22.809 | 30.375 | 75.993 | 70.259 | 70.259 |
| Azerbaijan | 19.521 | -22.809 | -3.288 | 42.331 | 58.437 | 58.437 |
| Bahrain | -146.875 | -22.809 | -169.684 | -124.066 | 0.000 | -0.000 |
| Botswana | -15.652 | -22.809 | -38.461 | 7.157 | 46.084 | 46.084 |
| Brazil | -120.687 | -22.809 | -143.496 | -97.878 | 9.197 | 9.197 |
| Bulgaria | 20.102 | -22.809 | -2.707 | 42.911 | 58.640 | 58.640 |
| Burkina Faso | -122.202 | -22.809 | -145.011 | -99.393 | 8.665 | 8.665 |
| Burundi | -35.091 | -22.809 | -57.900 | -12.282 | 39.257 | 39.257 |
| Chile | 28.513 | -22.809 | 5.703 | 51.322 | 61.594 | 61.594 |
| China | 39.895 | -22.809 | 17.086 | 62.705 | 65.592 | 65.592 |
| Columbia | -98.298 | -22.809 | -121.107 | -75.489 | 17.060 | 17.060 |
| Czech Republic | 40.722 | -22.809 | 17.913 | 63.531 | 65.882 | 65.882 |
| Denmark | 38.551 | -22.809 | 15.742 | 61.360 | 65.120 | 65.120 |
| Dominican | - | -22.809 | -123.666 | -78.048 | 16.161 | 16.161 |
| Republic | 100.857 | | | | | |
| Ecuador | -82.058 | -22.809 | -104.868 | -59.249 | 22.763 | 22.763 |
| Egypt | -69.567 | -22.809 | -92.376 | -46.758 | 27.150 | 27.150 |
| El Salvador | -86.836 | -22.809 | -109.646 | -64.027 | 21.085 | 21.085 |
| Estonia | 57.757 | -22.809 | 34.948 | 80.566 | 71.865 | 71.865 |
| France | 14.995 | -22.809 | -7.814 | 37.804 | 56.847 | 56.847 |
| Gambia, The | -95.399 | -22.809 | -118.208 | -72.590 | 18.078 | 18.078 |
| Georgia | 69.913 | -22.809 | 47.103 | 92.722 | 76.134 | 76.133 |
| Germany | 32.669 | -22.809 | 9.860 | 55.478 | 63.054 | 63.054 |
| Ghana | -129.214 | -22.809 | -152.024 | -106.405 | 6.202 | 6.202 |
| Greece | -8.468 | -22.809 | -31.278 | 14.341 | 48.607 | 48.607 |
| Guatemala | -114.803 | -22.809 | -137.612 | -91.993 | 11.264 | 11.264 |
| Guinea | -92.407 | -22.809 | -115.216 | -69.598 | 19.129 | 19.129 |
| Iceland | 80.522 | -22.809 | 57.713 | 103.332 | 79.860 | 79.860 |
| Indonesia | -126.339 | -22.809 | -149.148 | -103.530 | 7.212 | 7.212 |
| Israel | -54.918 | -22.809 | -77.728 | -32.109 | 32.294 | 32.294 |
| Italy | 4.905 | -22.809 | -17.904 | 27.714 | 53.304 | 53.304 |
| Jordan | -52.486 | -22.809 | -75.295 | -29.677 | 33.148 | 33.148 |
| Kazakhstan | 40.004 | -22.809 | 17.195 | 62.814 | 65.630 | 65.630 |
| Kenya | -68.807 | -22.809 | -91.616 | -45.998 | 27.417 | 27.417 |
| Latvia | 39.339 | -22.809 | 16.530 | 62.148 | 65.396 | 65.396 |
| Lebanon | -2.511 | -22.809 | -25.320 | 20.299 | 50.699 | 50.699 |
| Lesotho | 14.681 | -22.809 | -8.128 | 37.490 | 56.737 | 56.737 |
| Madagascar | -72.217 | -22.809 | -95.027 | -49.408 | 26.219 | 26.219 |
| Malawi | -82.106 | -22.809 | -104.915 | -59.297 | 22.746 | 22.746 |
| Malaysia | -115.477 | -22.809 | -138.286 | -92.668 | 11.027 | 11.027 |
| Mali | -104.631 | -22.809 | -127.440 | -81.822 | 14.836 | 14.836 |
| Malta | -29.098 | -22.809 | -51.907 | -6.289 | 41.362 | 41.362 |
| Mauritius | -73.467 | -22.809 | -96.276 | -50.658 | 25.780 | 25.780 |
| Mexico | -66.103 | -22.809 | -88.912 | -43.294 | 28.366 | 28.366 |
| Moldova | 28.621 | -22.809 | 5.812 | 51.430 | 61.632 | 61.632 |
| Mongolia | 93.886 | -22.809 | 71.077 | 116.695 | 84.553 | 84.553 |
| Morocco | -42.498 | -22.809 | -65.308 | -19.689 | 36.656 | 36.656 |
| Mozambique | -61.172 | -22.809 | -83.981 | -38.363 | 30.098 | 30.098 |
| Namibia | -42.455 | -22.809 | -65.264 | -19.646 | 36.671 | 36.671 |
| Nepal | -8.864 | -22.809 | -31.674 | 13.945 | 48.468 | 48.468 |
| Netherlands | 30.211 | -22.809 | 7.401 | 53.020 | 62.191 | 62.191 |
| Norway | 59.262 | -22.809 | 36.452 | 82.071 | 72.393 | 72.393 |
| Oman | -83.642 | -22.809 | -106.451 | -60.833 | 22.207 | 22.207 |
| Pakistan | -54.795 | -22.809 | -77.604 | -31.986 | 32.338 | 32.338 |
| Paraguay | -92.331 | -22.809 | -115.141 | -69.522 | 19.155 | 19.155 |
| Peru | -59.173 | -22.809 | -81.982 | -36.364 | 30.800 | 30.800 |
| Philippines | -112.126 | -22.809 | -134.935 | -89.317 | 12.204 | 12.204 |
| Poland | 39.087 | -22.809 | 16.278 | 61.896 | 65.308 | 65.308 |
| Portugal | -30.500 | -22.809 | -53.309 | -7.690 | 40.870 | 40.870 |
| Romania | 37.092 | -22.809 | 14.283 | 59.901 | 64.607 | 64.607 |
| Russian Federation | 137.872 | -22.809 | 115.062 | 160.681 | 100.000 | 100.000 |
| Rwanda | -14.241 | -22.809 | -37.050 | 8.568 | 46.580 | 46.580 |
| Saudi Arabia | -72.977 | -22.809 | -95.786 | -50.168 | 25.952 | 25.952 |
| Spain | -7.513 | -22.809 | -30.322 | 15.296 | 48.942 | 48.942 |
| Sri Lanka | -121.383 | -22.809 | -144.192 | -98.574 | 8.953 | 8.953 |
| Sweden | 88.149 | -22.809 | 65.340 | 110.958 | 82.538 | 82.538 |
| Switzerland | 56.394 | -22.809 | 33.585 | 79.203 | 71.386 | 71.386 |
| Thailand | -115.024 | -22.809 | -137.833 | -92.215 | 11.186 | 11.186 |
| Turkey | 11.822 | -22.809 | -10.987 | 34.631 | 55.733 | 55.733 |
| Uganda | -52.025 | -22.809 | -74.834 | -29.216 | 33.310 | 33.310 |
| Ukraine | 18.744 | -22.809 | -4.065 | 41.554 | 58.164 | 58.164 |
| United Kingdom | 37.663 | -22.809 | 14.854 | 60.473 | 64.808 | 64.808 |
| United States | 38.354 | -22.809 | 15.545 | 61.163 | 65.050 | 65.050 |
| Zambia | -76.079 | -22.809 | -98.888 | -53.270 | 24.863 | 24.863 |

TABLE 7: GACI calculated scores and rankings

| Country | GACI Ranking | GACI Scores 2019 |
|---|---|---|
| United States | 1 | 80.676 |
| Switzerland | 2 | 79.803 |
| Sweden | 3 | 79.019 |
| Germany | 4 | 78.481 |
| Netherlands | 5 | 78.018 |
| United Kingdom | 6 | 77.738 |
| Denmark | 7 | 77.075 |
| Norway | 8 | 75.702 |
| France | 9 | 75.033 |
| Austria | 10 | 74.268 |
| Iceland | 11 | 74.094 |
| China | 12 | 73.242 |
| Czech Republic | 13 | 69.100 |
| Estonia | 14 | 68.785 |
| Russian Federation | 15 | 68.757 |
| Australia | 16 | 68.512 |
| Italy | 17 | 68.312 |
| Chile | 18 | 68.267 |
| Poland | 19 | 67.284 |
| Malaysia | 20 | 66.547 |
| Israel | 21 | 66.102 |
| Spain | 22 | 64.532 |
| Latvia | 23 | 64.313 |
| Romania | 24 | 63.520 |
| Bulgaria | 25 | 63.123 |
| Kazakhstan | 26 | 63.017 |
| Saudi Arabia | 27 | 62.596 |
| Thailand | 28 | 61.712 |
| Azerbaijan | 29 | 61.622 |
| Turkey | 30 | 61.314 |
| Georgia | 31 | 60.526 |
| Portugal | 32 | 60.467 |
| Armenia | 33 | 60.191 |
| Indonesia | 34 | 59.131 |
| Bahrain | 35 | 58.509 |
| Malta | 36 | 57.768 |
| Mexico | 37 | 57.486 |
| Columbia | 38 | 56.510 |
| Mauritius | 39 | 56.279 |
| Moldova | 40 | 56.225 |
| Oman | 41 | 55.961 |
| Ukraine | 42 | 55.923 |
| Philippines | 43 | 55.758 |
| Brazil | 44 | 55.358 |
| Peru | 45 | 54.128 |
| Mongolia | 46 | 53.819 |
| Greece | 47 | 52.139 |
| Jordan | 48 | 52.071 |
| Dominican Republic | 49 | 51.986 |
| Sri Lanka | 50 | 51.832 |
| Morocco | 51 | 50.653 |
| Ecuador | 52 | 48.541 |
| Egypt | 53 | 48.142 |
| Argentina | 54 | 48.087 |
| Albania | 55 | 47.963 |
| Guatemala | 56 | 47.773 |
| Botswana | 57 | 47.138 |
| Paraguay | 58 | 46.875 |
| Namibia | 59 | 46.727 |
| Kenya | 60 | 46.554 |
| Ghana | 61 | 46.435 |
| Lebanon | 62 | 46.373 |
| El Salvador | 63 | 46.131 |
| Pakistan | 64 | 45.022 |
| Rwanda | 65 | 44.741 |
| Nepal | 66 | 43.736 |
| Lesotho | 67 | 43.313 |
| Uganda | 68 | 41.883 |
| Zambia | 69 | 41.828 |
| Guinea | 70 | 40.408 |
| Malawi | 71 | 39.375 |
| Mali | 72 | 39.248 |
| Burkina Faso | 73 | 39.073 |
| Gambia, The | 74 | 38.797 |
| Madagascar | 75 | 37.506 |
| Burundi | 76 | 34.088 |
| Angola | 77 | 33.496 |
| Mozambique | 78 | 32.623 |

# FIGURES

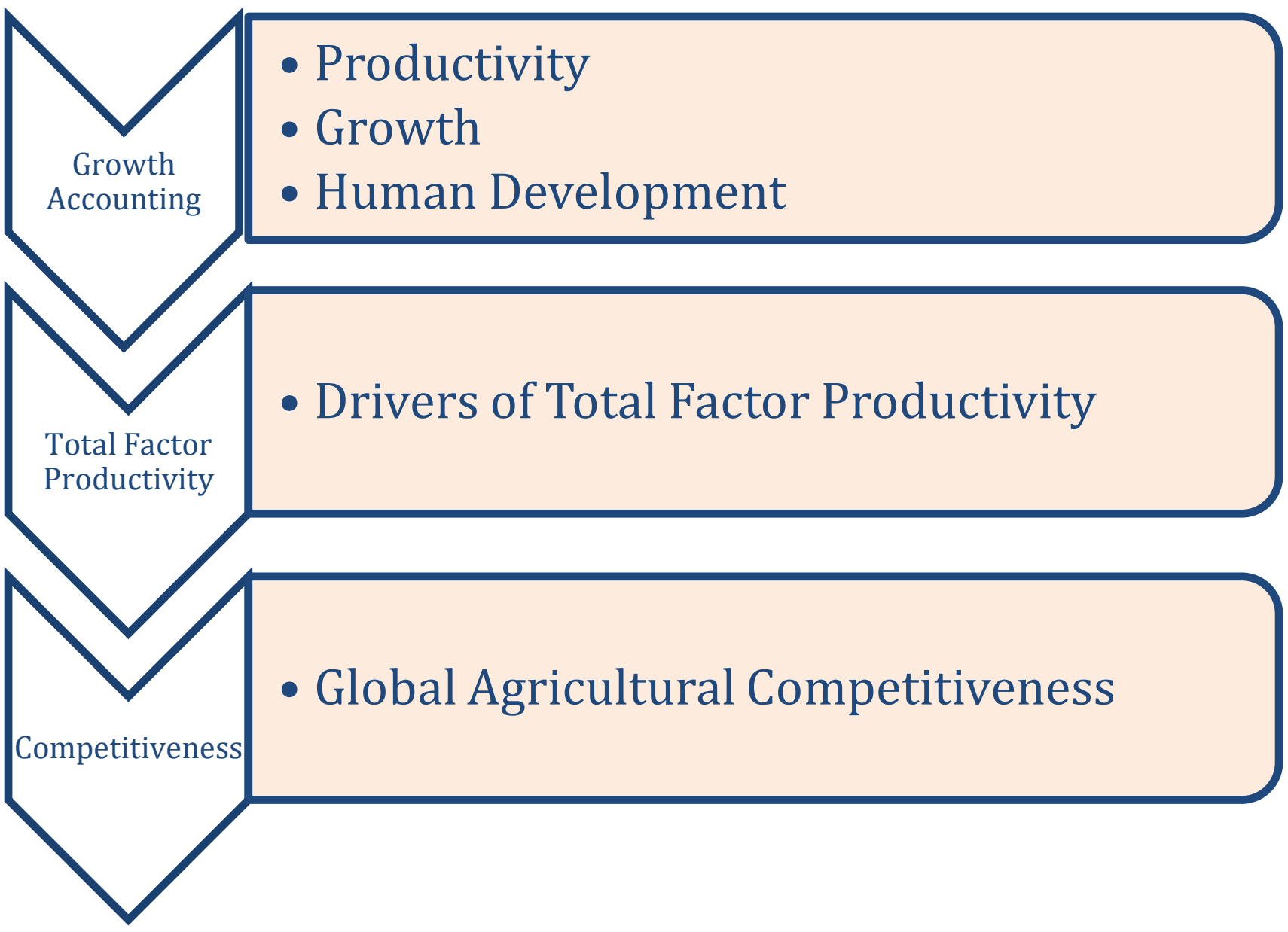

FIGURE 1: Theoretical framework for developing a global agricultural competitiveness index.

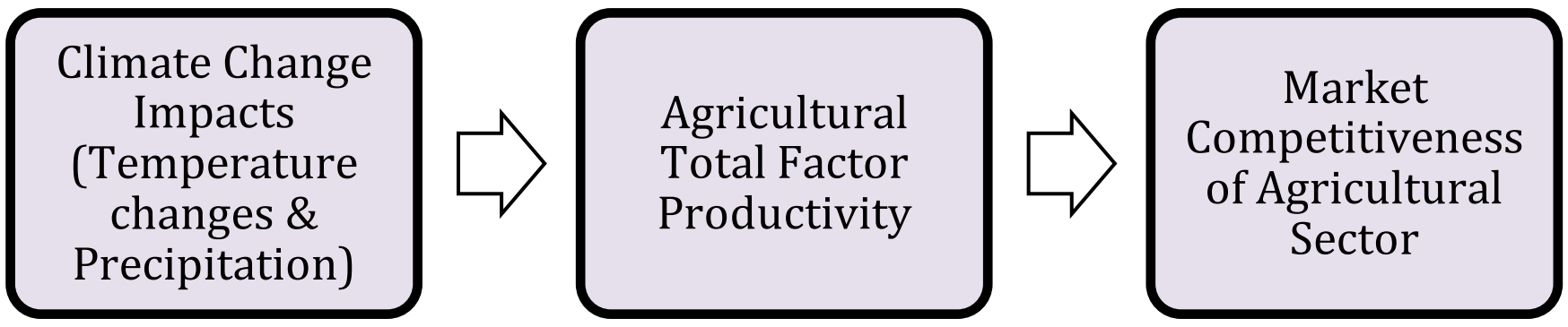

FIGURE 2: Conceptual framework for construction of pillar 14 (climate change impact evaluation)

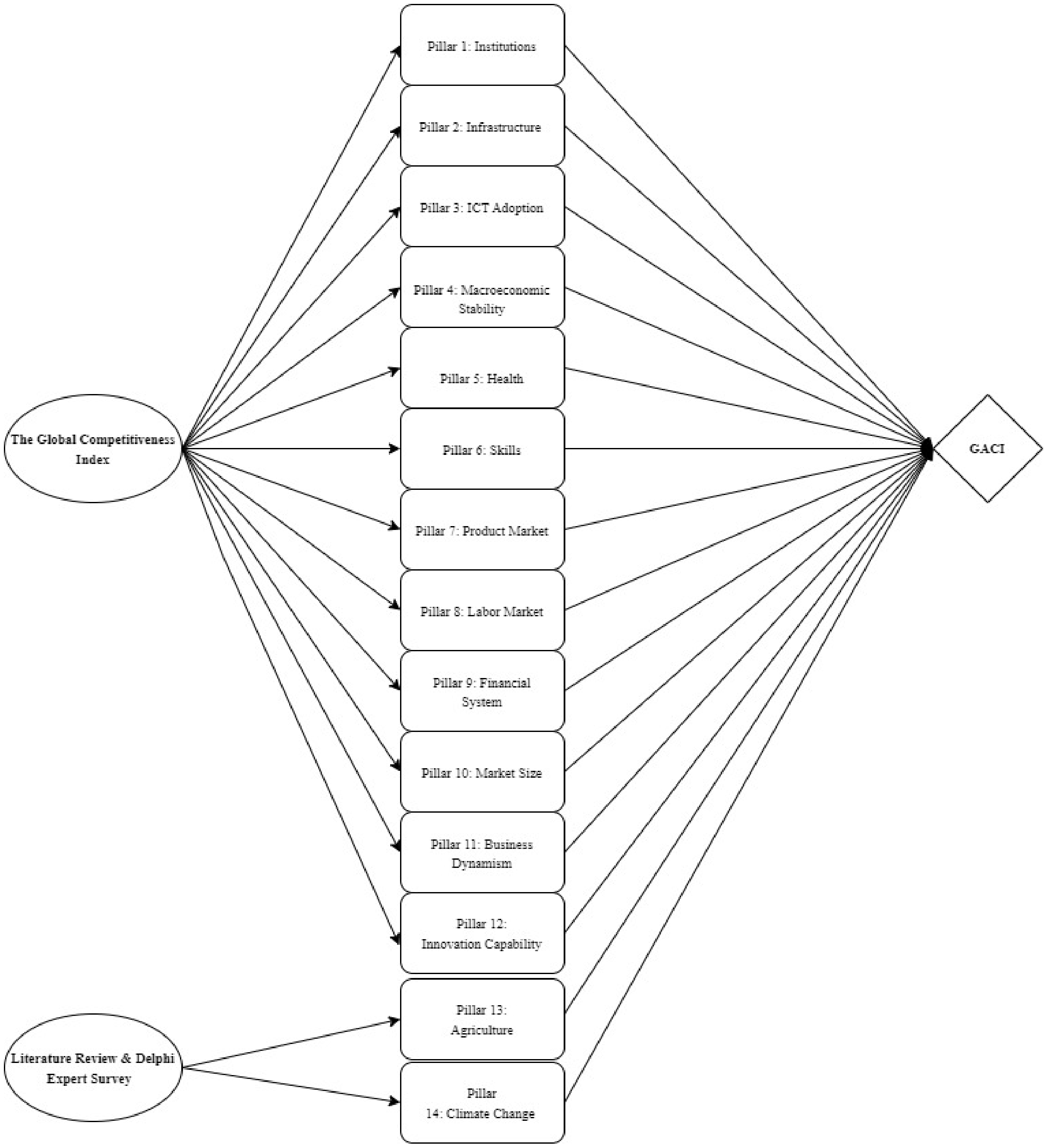

Source: (Zia et al., 2022)

FIGURE 3. Framework designed for the global agricultural competitiveness assessment.

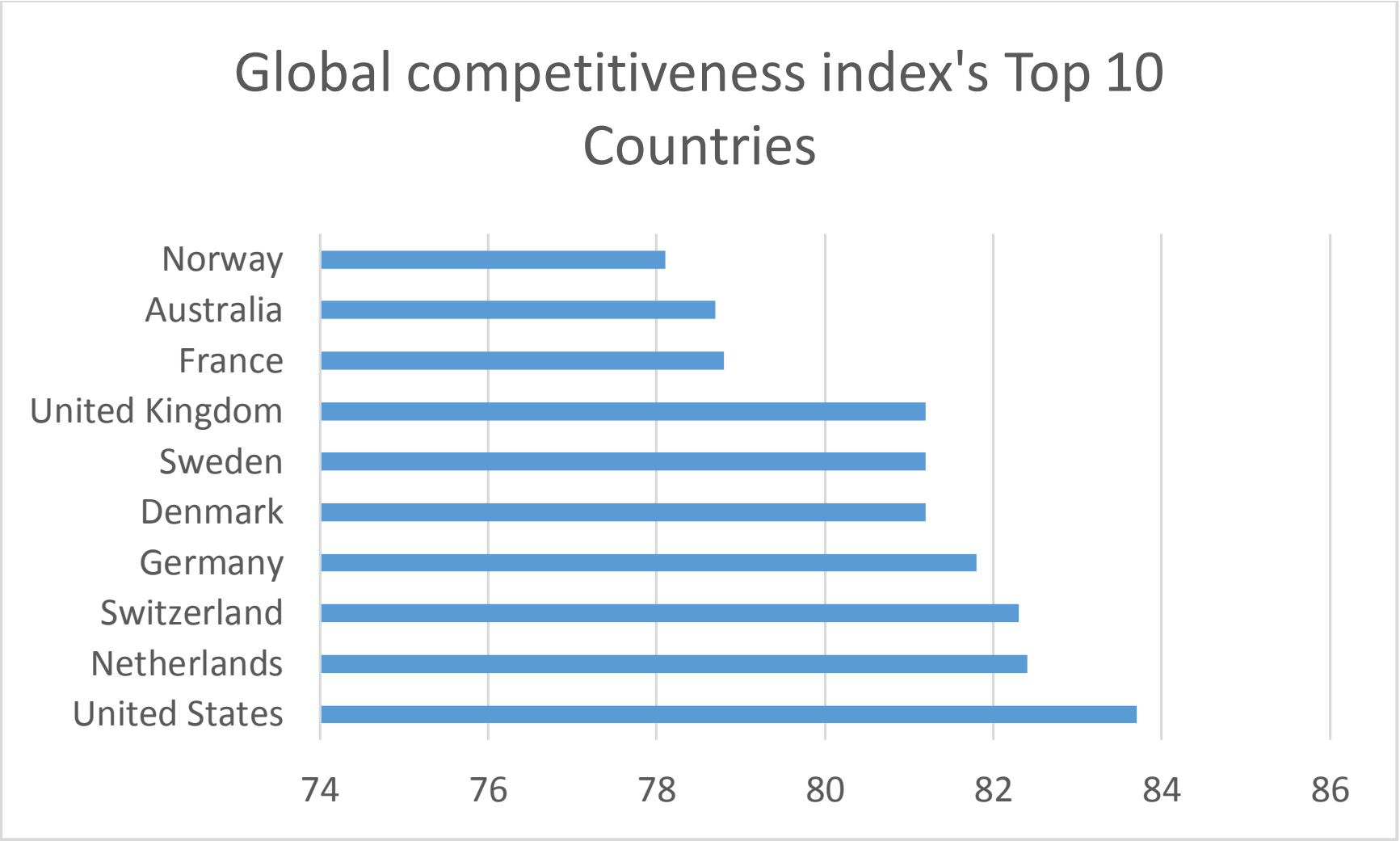

Source: Author's reranking of global competitiveness index (GCI) scores within the selected countries.

FIGURE 4: Top 10 countries in the global competitiveness index (within the 78 selected countries).

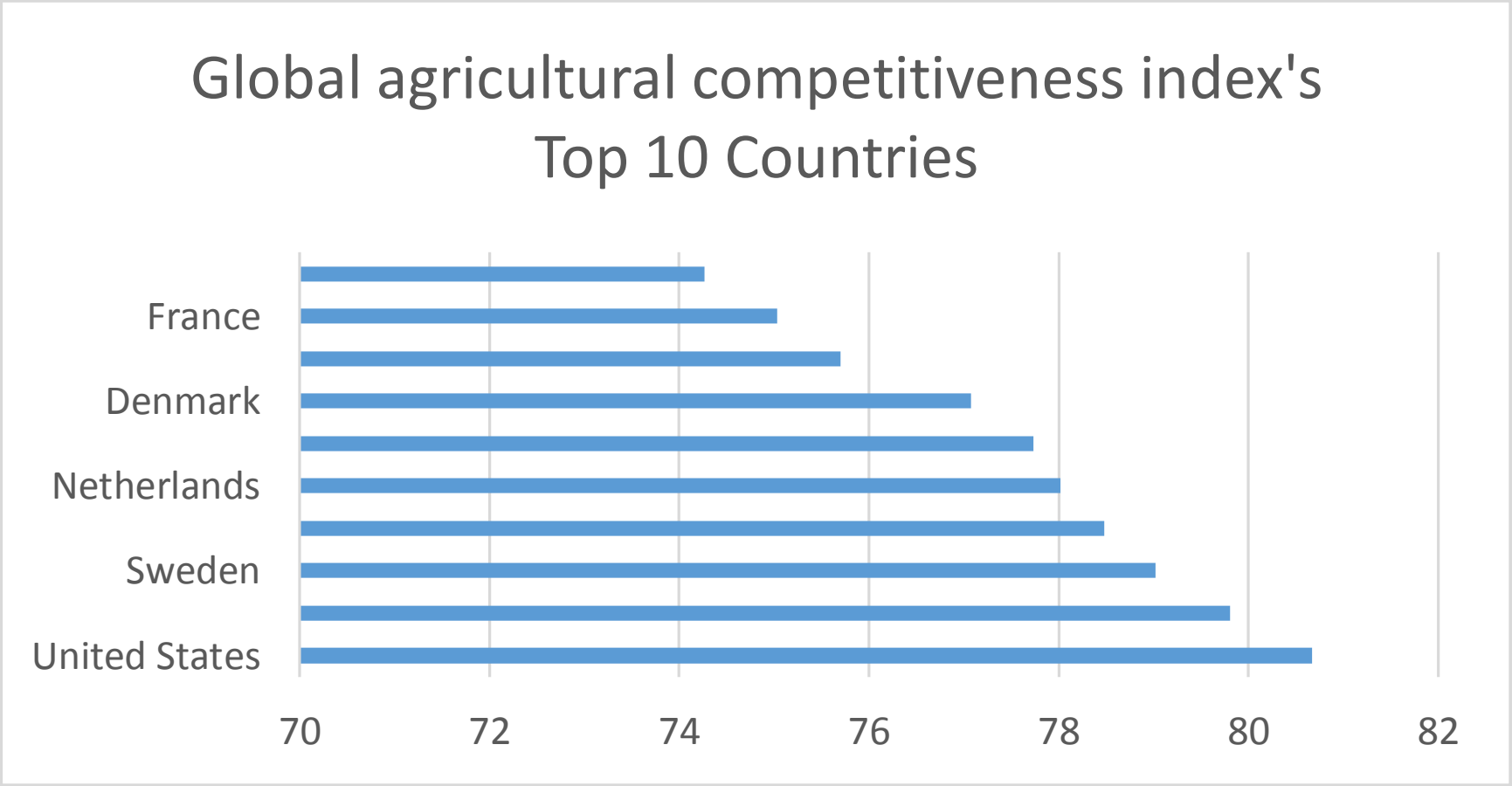

Source: Author calculations.

FIGURE 5: Top 10 Countries in the global agricultural competitiveness index.

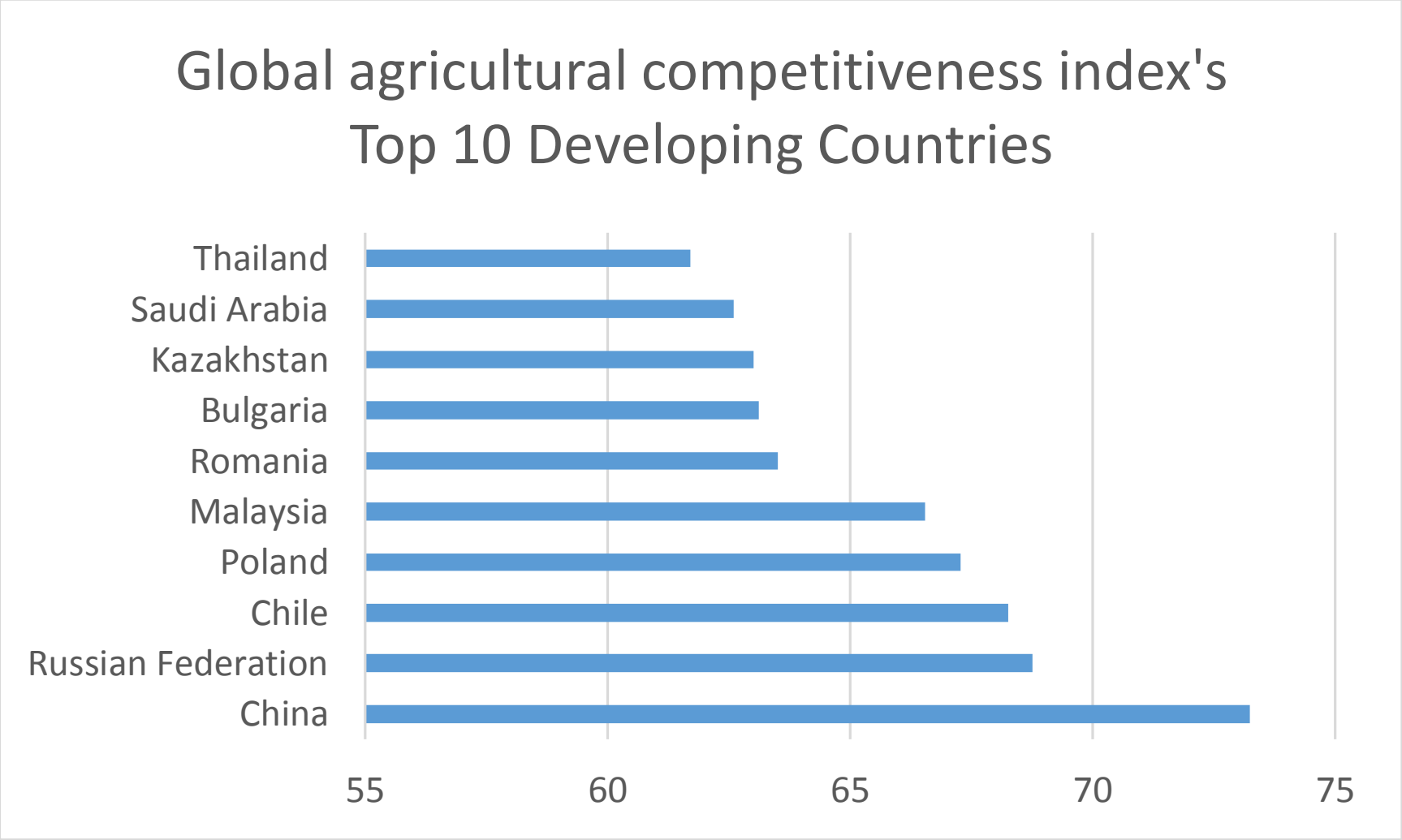

Source: Author calculations.

FIGURE 6: Top 10 Developing Countries in the global agricultural competitiveness index.

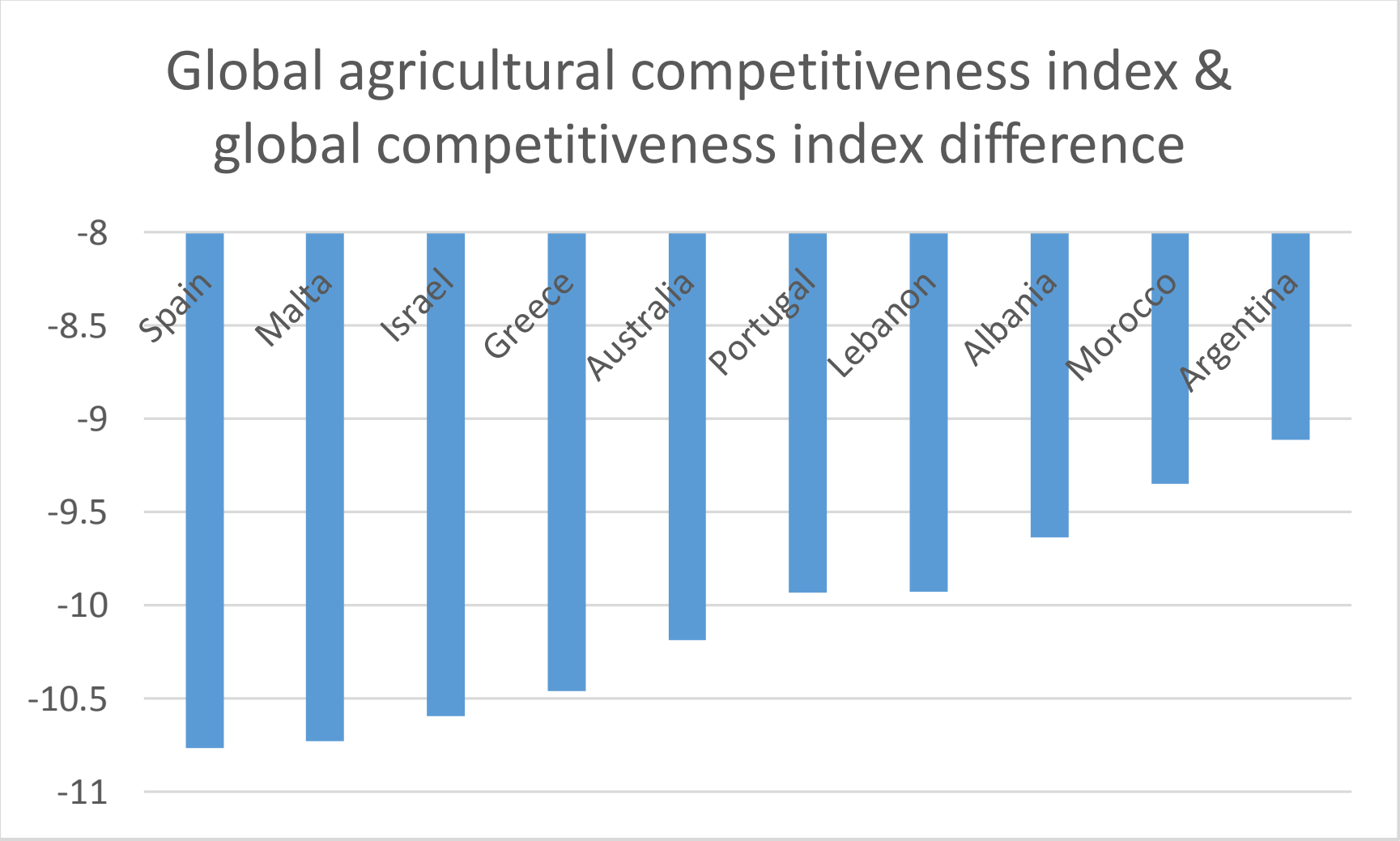

FIGURE 7: Countries with maximum differences between the global agricultural competitiveness index and the global competitiveness index.

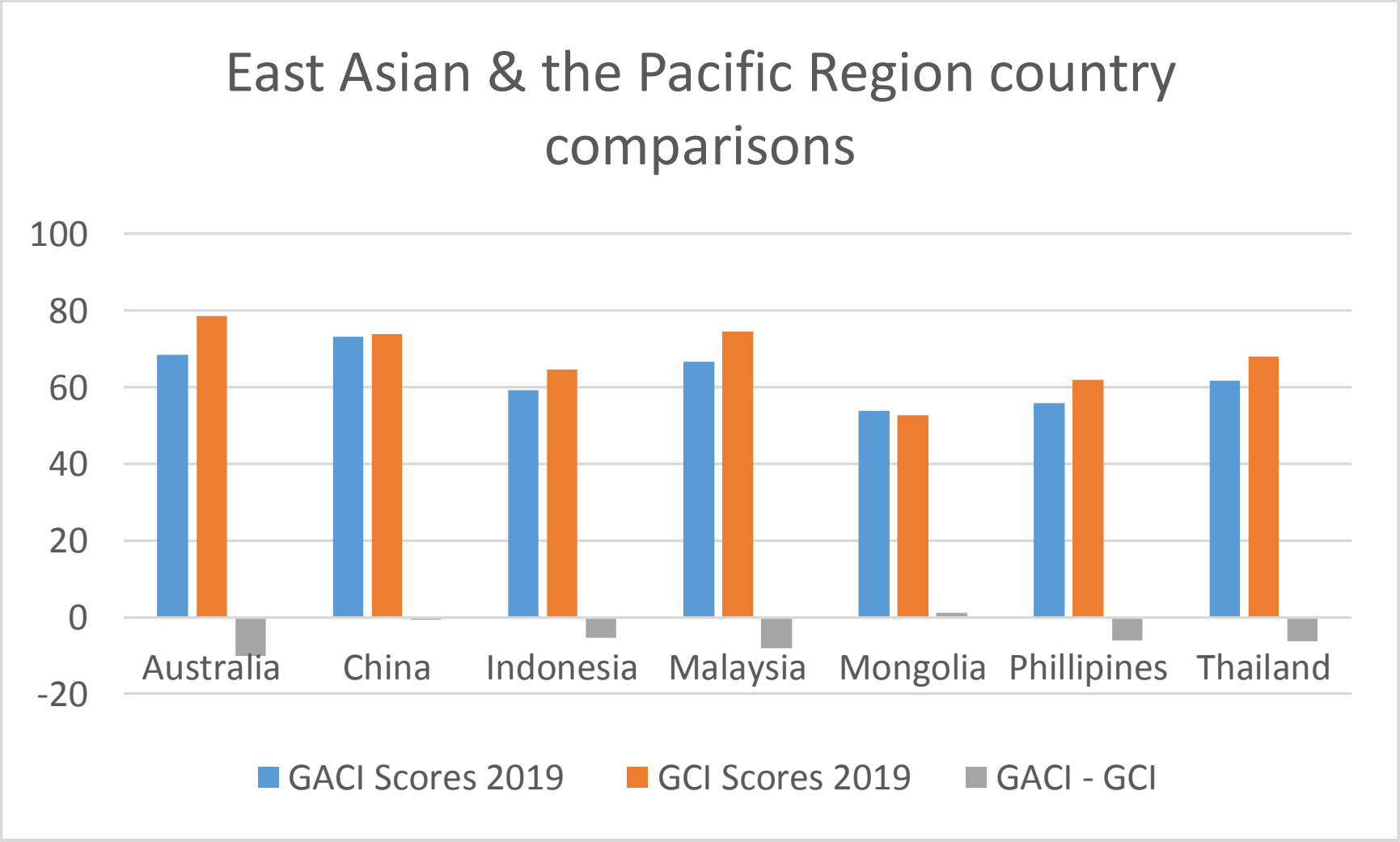

Source: Author calculations.

FIGURE 8: Comparison between the East Asian and Pacific region countries.

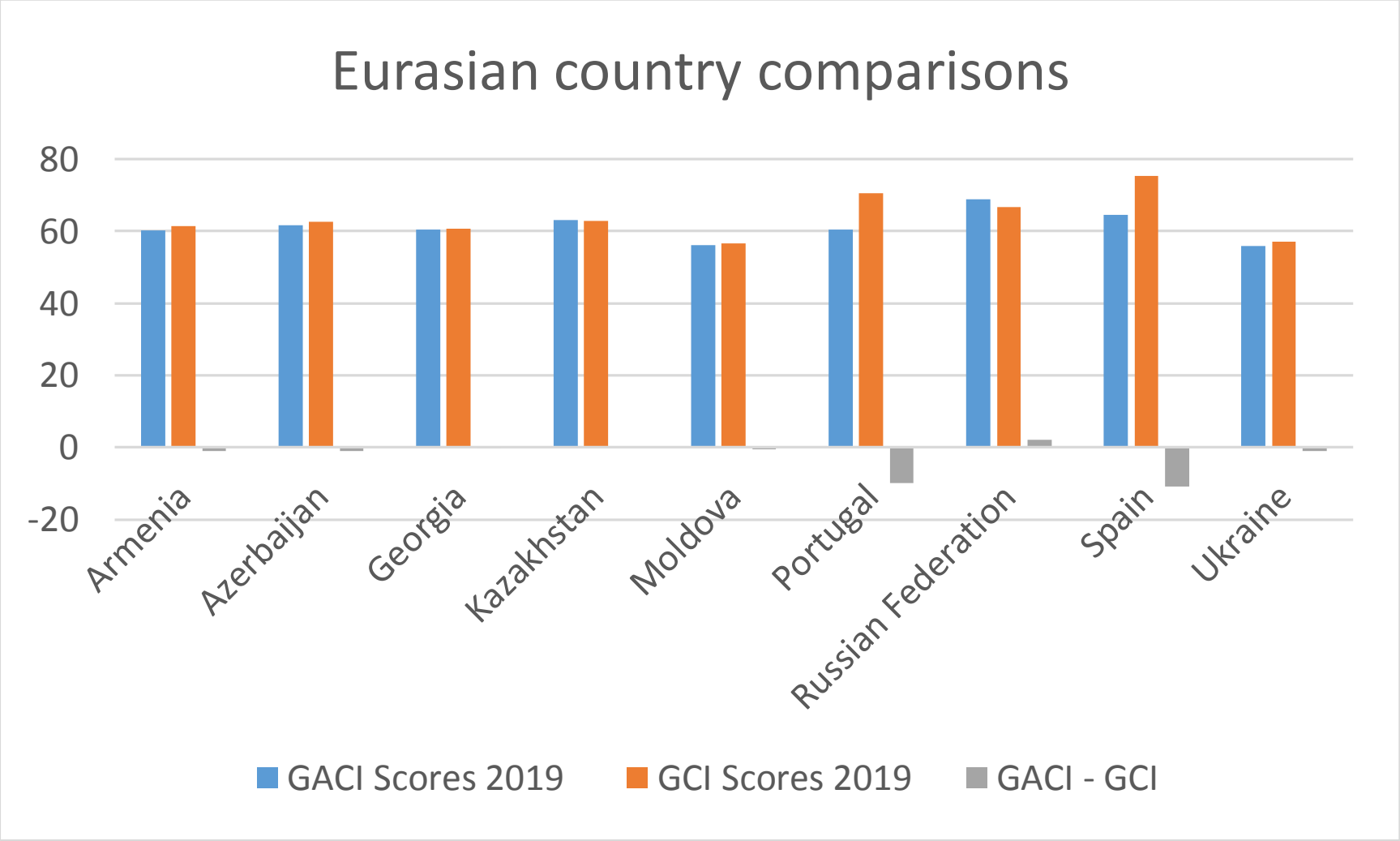

Source: Author calculations.

FIGURE 9: Comparison between Eurasian countries

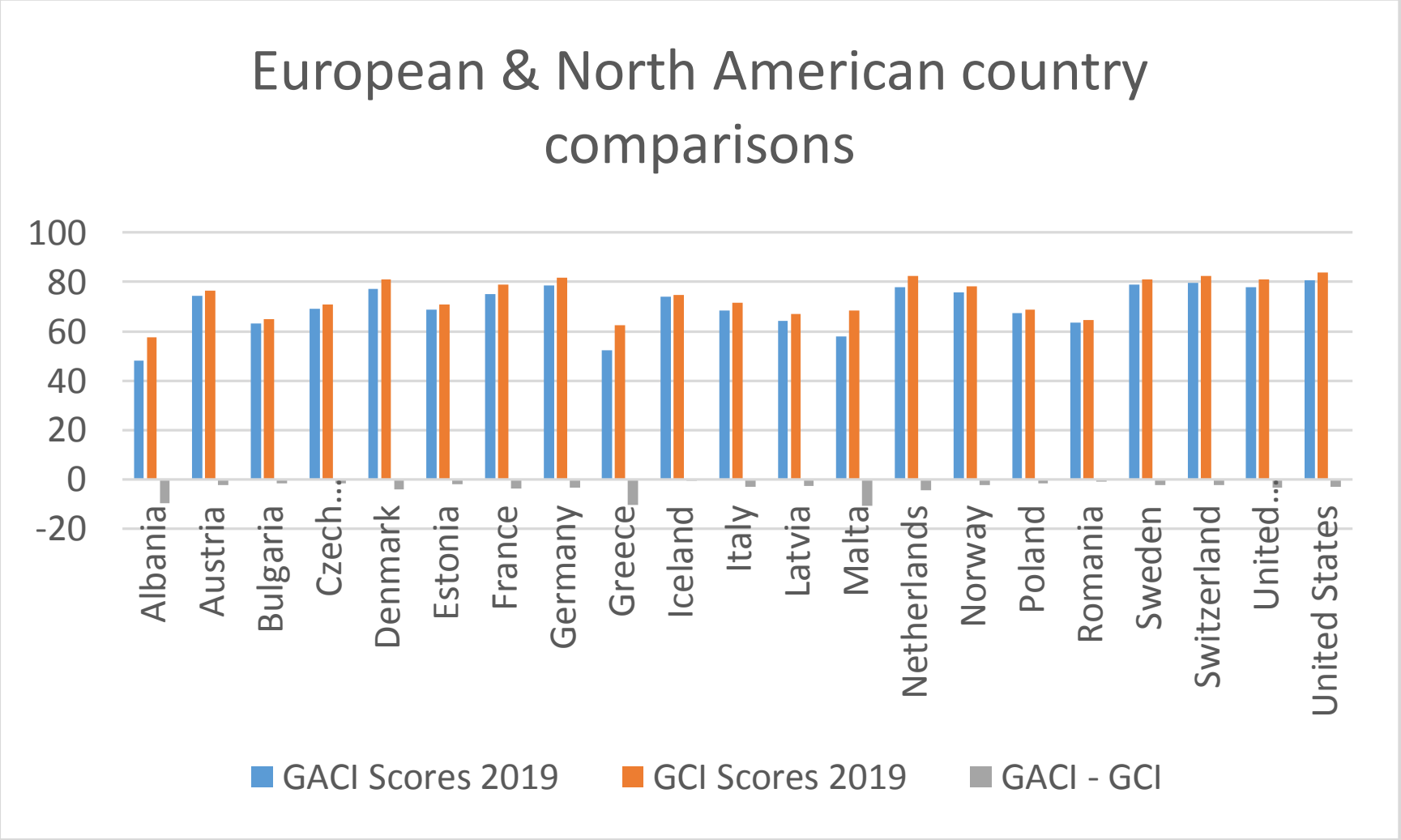

Source: Author calculations.

FIGURE 10: Comparison between European and North American countries

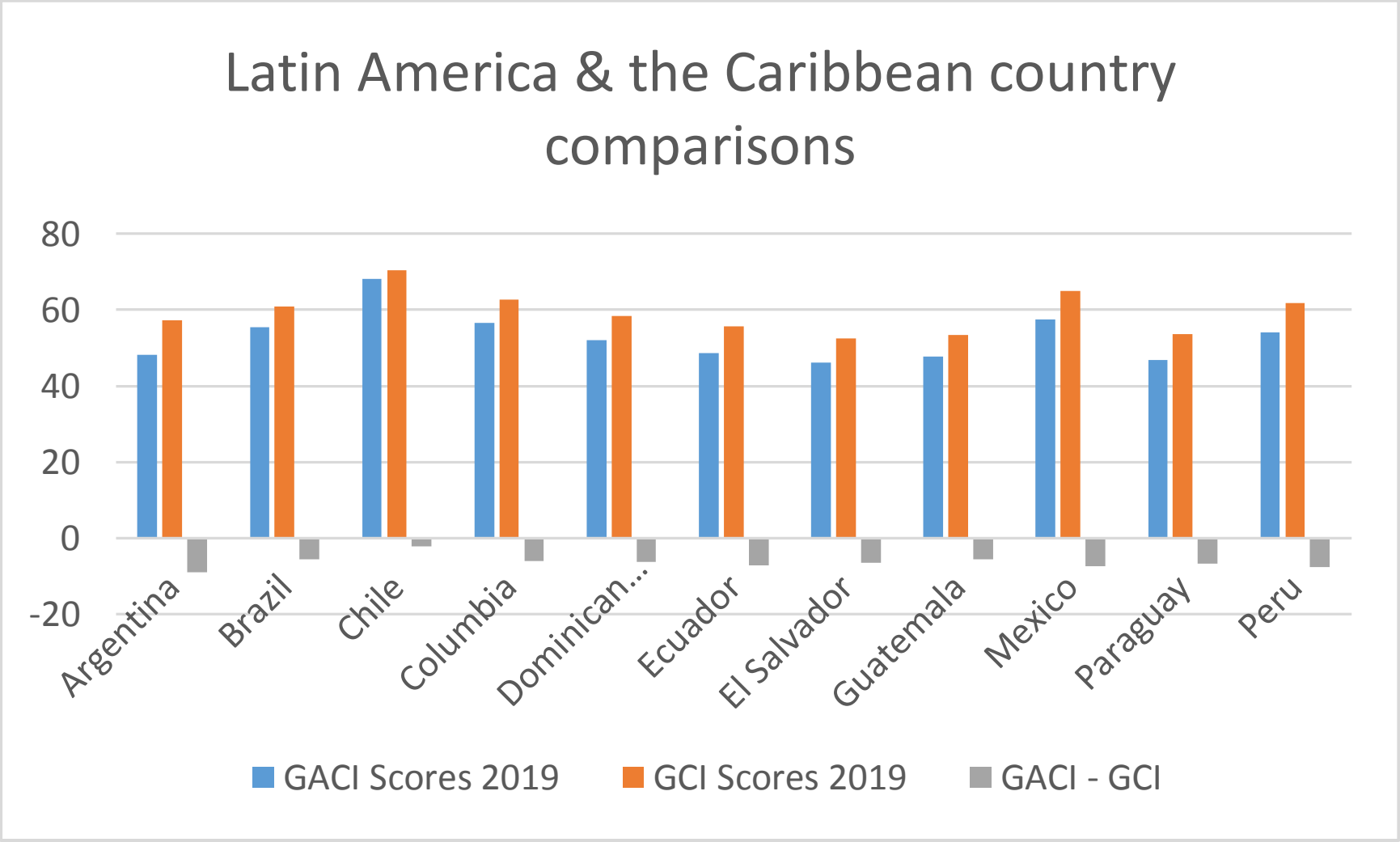

Source: Author calculations.

FIGURE 11: Comparison between Latin American and Caribbean countries

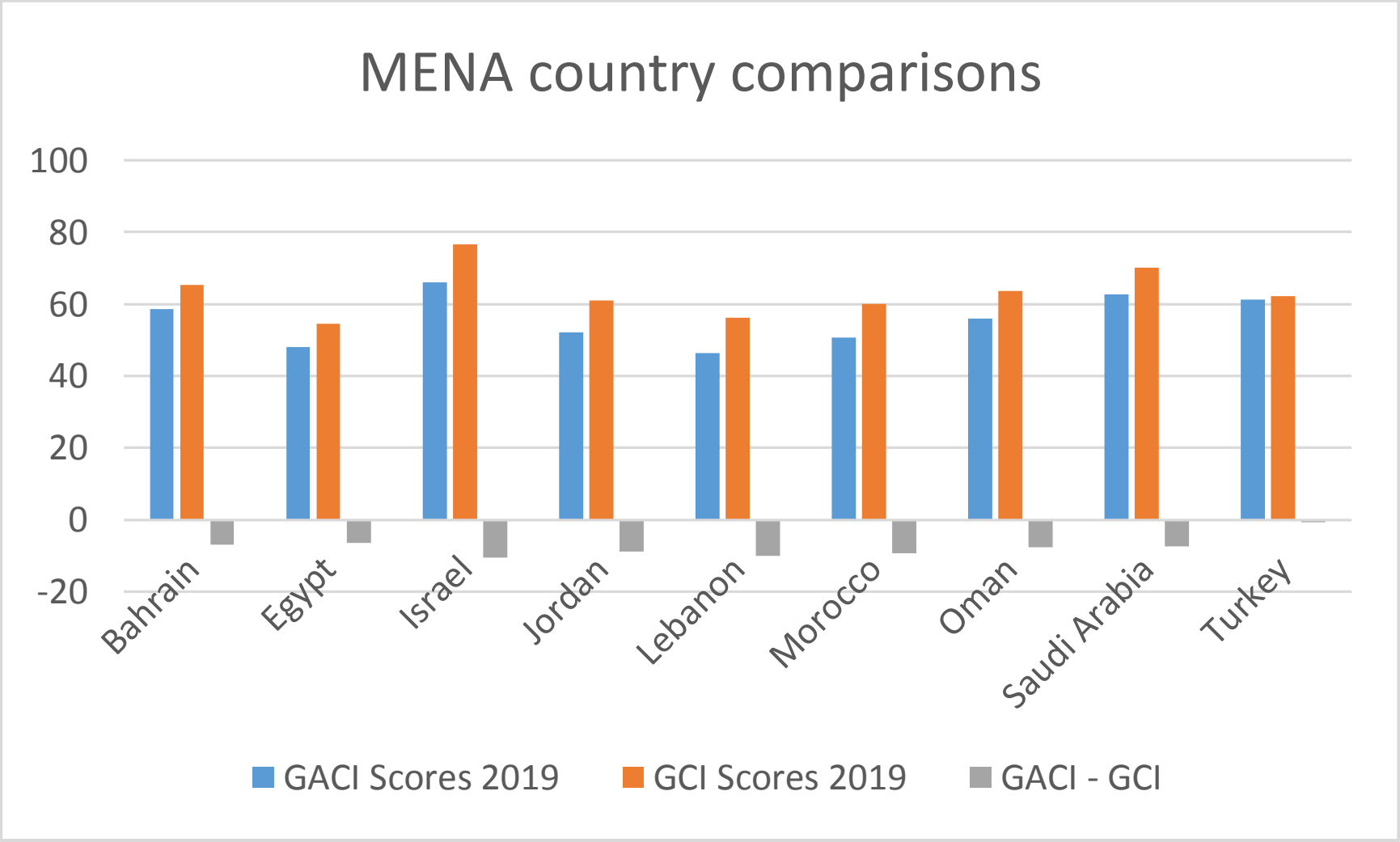

Source: Author calculations.

FIGURE 12: Comparison between Middle East and North African countries

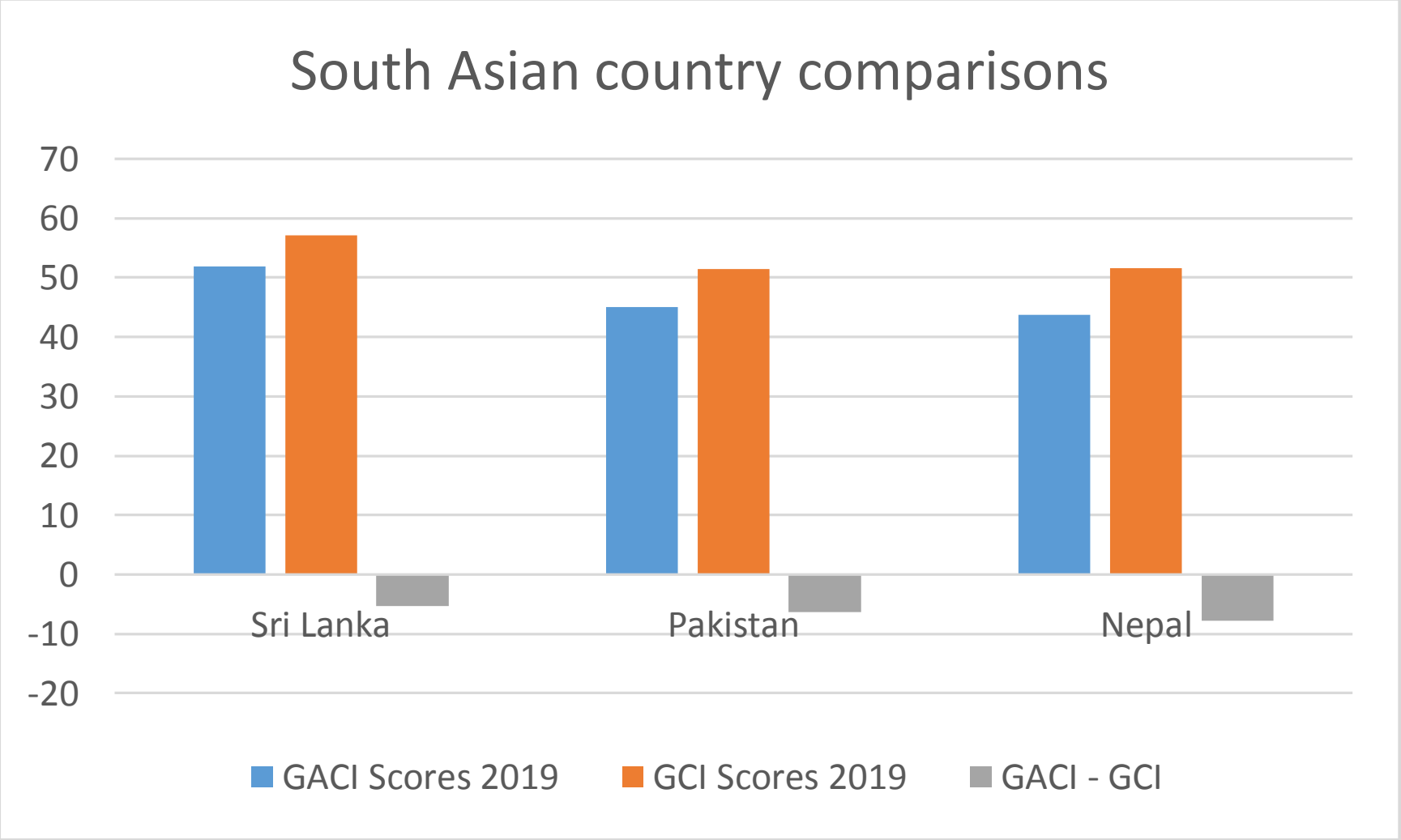

Source: Author calculations.

FIGURE 13: Comparison between South Asian countries.

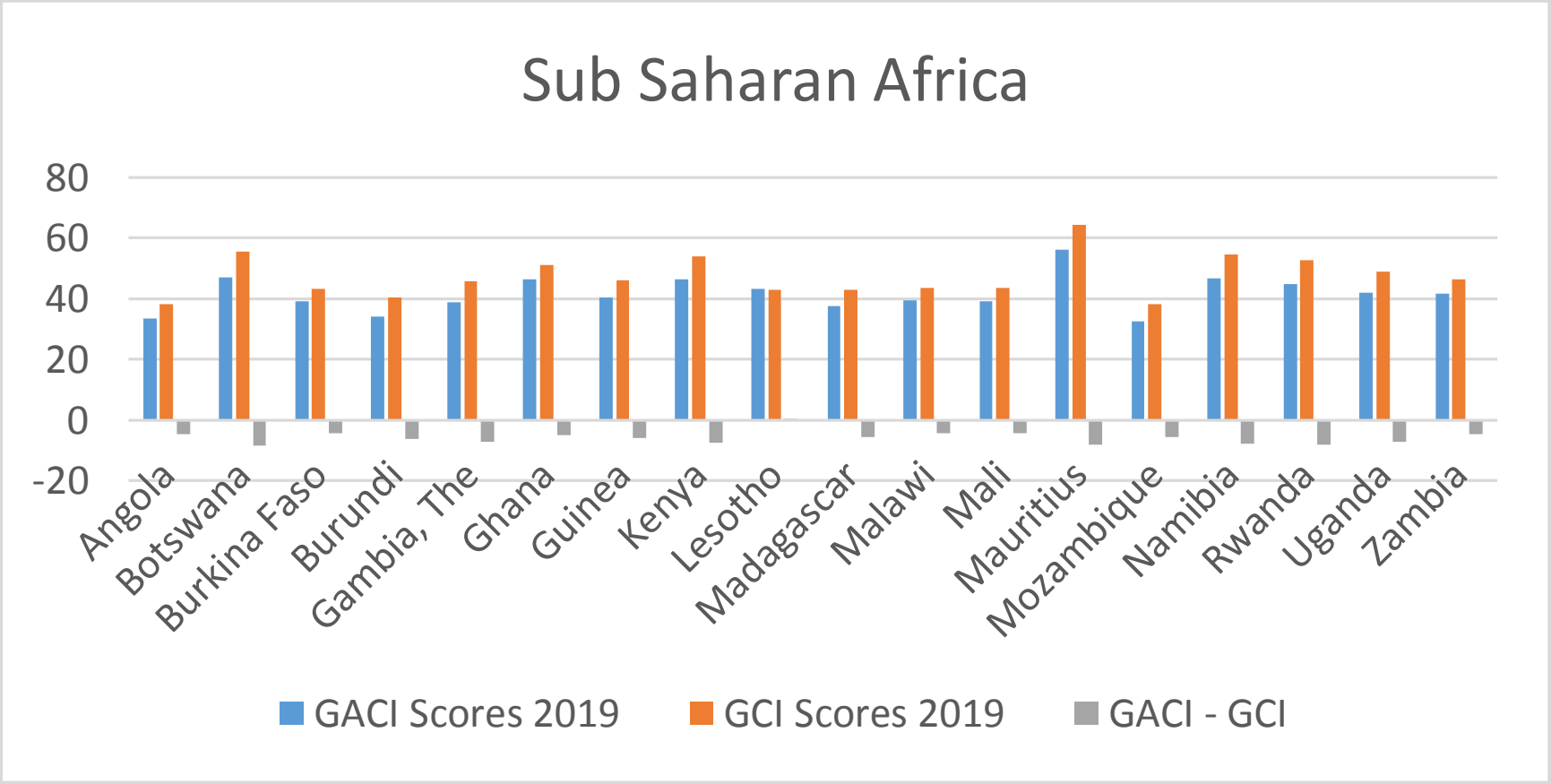

Source: Author calculations.

FIGURE 14: Comparison between Sub-Saharan Africa countries.

## Appendix Table 1: Pillars of Global Competitiveness Index (GCI)

| **ENABLING ENVIRONMENT (not used in calculation)** | | |
|---|---|---|
| PILLAR 1<br>Institutions | A. Security | 1.01 Organized crime<br>1.02 Homicide rate<br>1.03 Terrorism incidence<br>1.04 Reliability of police services |
| | B. Social Capital | 1.05Social capital |
| | C. Checks & Balances | 1.06 Budget transparency<br>1.07 Judicial independence<br>1.08 Efficiency of legal framework in challenging regulations<br>1.09Freedom of the press |
| | D. Public Sector Performance | 1.10 Burden of government regulation<br>1.11 Efficiency of legal framework in settling disputes<br>1.12-Participation |
| | E. Transparency | 1.13Incidence of corruption |
| | F. Property Rights | 1.14 Property rights<br>1.15 Intellectual property protection<br>1.16Quality of land administration |
| | G. Corporate Governance | 1.17 Strength of auditing and accounting standards<br>1.18 Conflict of interest regulation<br>1.19Shareholder governance |
| | | I. Government adaptability<br>1.20 Government ensuring policy stability<br>1.21 Government's responsiveness to change |
| | H. Future Orientation of the Governments | 1.22 Legal framework's adaptability to digital business models<br>1.23 Government long-term vision<br>II. Commitment to sustainability<br>1.24 Energy efficiency regulation<br>1.25 Renewable energy regulation<br>1.26 Environment-related treaties in force |
| PILLAR 2<br>Infrastructure | A. Transport Infrastructure | I. Road<br>2.01 Road connectivity<br>2.02 Quality of road infrastructure<br>II. Railroad<br>2.03 Railroad density<br>2.04 Efficiency of train services<br>III. Air<br>2.05 Airport connectivity<br>2.06 Efficiency of air transport services<br>IV. Sea<br>2.07 Liner shipping connectivity3<br>2.08Efficiency of seaport services |
| | B. Utility Infrastructure | I. Electricity<br>2.09 Electricity access<br>2.10 Electricity supply quality<br>II. Water<br>2.11 Exposure to unsafe drinking water<br>2.12 Reliability of water supply |
| PILLAR 3<br>ICT Adoption | | 3.01 Mobile-cellular telephone subscriptions<br>3.02 Mobile-broadband subscriptions<br>3.03 Fixed-broadband internet subscriptions<br>3.04 Fiber internet subscriptions<br>3.05 Internet users |
| PILLAR 4<br>Macroeconomic Stability | | 4.01 Inflation<br>4.02 Debt dynamics |
| **HUMAN CAPITAL (not used in calculation)** | | |
| PILLAR 5<br>Health | | 5.01 Healthy life expectancy |
| PILLAR 6<br>Skills | A. Current Workforce | I. Education of current workforce<br>6.01 Mean years of schooling<br>II. Skills of current workforce<br>6.02 Extent of staff training<br>6.03 Quality of vocational training<br>6.04 Skillset of graduates<br>6.05 Digital skills among active population<br>6.06 Ease of finding skilled employees |
| | B. Future Workforce | I. Education of future workforce<br>6.07 School life expectancy<br>II. Skills of future workforce<br>6.08 Critical thinking in teaching<br>6.09 Pupil-to-teacher ratio in primary education |
| **MARKETS (not used in calculation)** | | |
| PILLAR 7<br>Product Market | A. Domestic Market Competition | 7.01 Distortive effect of taxes and subsidies on competition<br>7.02 Extent of market dominance<br>7.03Competition in services |
| | B. Trade Openness | 7.04 Prevalence of nontariff barriers<br>7.05 Trade tariffs<br>7.06 Complexity of tariffs<br>7.07 Border clearance efficiency |
| PILLAR 8<br>Labor Market | A. Flexibility | 8.01 Redundancy costs<br>8.02 Hiring and firing practices<br>8.03 Cooperation in labor-employer relations<br>8.04 Flexibility of wage determination<br>8.05 Active labor market policies<br>8.06 Workers' rights<br>8.07 Ease of hiring foreign labor<br>8.08 Internal labor mobility |
| | B. Meritocracy and Incentivization | 8.09 Reliance on professional management<br>8.10 Pay and productivity<br>8.11 Ratio of wage and salaried female workers to male workers<br>8.12 Labor tax rate |
| PILLAR 9<br>Financial System | A. Depth | 9.01 Domestic credit to private sector<br>9.02 Financing of SMEs<br>9.03 Venture capital availability<br>9.04 Market capitalization<br>9.05Insurance premium |
| | | 9.06 Soundness of banks |
| | B. Stability | 9.07 nonperforming loans<br>9.08 Credit gap<br>9.09 Banks' regulatory capital ratio |
| PILLAR 10<br>Market Size | | 10.01 Gross domestic product<br>10.02 Imports of goods and services |
| **INNOVATION ECOSYSTEM (not used in calculation)** | | |
| PILLAR 11<br>Business Dynamism | A. Administrative Requirements | 11.01 Cost of starting a business<br>11.02 Time to start a business<br>11.03 Insolvency recovery rate<br>11.04 Insolvency regulatory framework |
| | B. Entrepreneurial culture | 11.05 Attitudes toward entrepreneurial risk<br>11.06 Willingness to delegate authority<br>11.07 Growth of innovative companies<br>11.08 Companies embracing disruptive ideas |
| PILLAR 12<br>Innovation Capability | A. Diversity and collaboration | 12.01 Diversity of workforce<br>12.02 State of cluster development<br>12.03 International conventions<br>12.04 Multistakeholder collaboration |
| | B. Research and development | 12.05 Scientific publications<br>12.06 Patent applications<br>12.07 R&D expenditures<br>12.08 Research institutions prominence index |
| | A. Commercialization | 12.09 Buyer sophistication<br>12.10 Trademark applications |

**Appendix Table 2: List of Pillars for Global Agricultural Competitiveness Index (GACI)**

| Pillar/ Country | 1 | 2 | 3 | 4 | 5 | 6 | 7 | 8 | 9 | 10 | 11 | 12 | 13 | 14 |
|---|---|---|---|---|---|---|---|---|---|---|---|---|---|---|
| Albania | 51.88 | 57.70 | 52.90 | 70 | 85.90 | 68.96 | 54.38 | 65.26 | 53.30 | 39.60 | 61.80 | 29.74 | 31.66 | -51.58 |
| Angola | 37.62 | 40.19 | 30.50 | 40.60 | 46.90 | 29.06 | 37.74 | 46.82 | 38.42 | 53.90 | 36.75 | 18.82 | 34.04 | -22.41 |
| Argentina | 49.85 | 68.288 | 58 | 33.90 | 83.80 | 72.27 | 46.95 | 51.83 | 52.86 | 68.60 | 58.29 | 41.74 | 34.79 | -47.94 |
| Armenia | 56.25 | 69.41 | 62 | 75 | 80.70 | 66.78 | 59.08 | 66.44 | 60.14 | 37.50 | 62.55 | 39.40 | 43.11 | 64.32 |
| Australia | 72.94 | 79.16 | 73.60 | 100 | 94.90 | 80.56 | 71.39 | 69.07 | 85.90 | 72.60 | 75.30 | 69.55 | 42.12 | -27.92 |
| Austria | 73.55 | 89.05 | 65.60 | 100 | 95.10 | 79.36 | 66.09 | 67.16 | 74.98 | 64.60 | 69.35 | 74.47 | 50.19 | 70.26 |
| Azerbaijan | 58.47 | 77.37 | 55.10 | 70.05 | 68.90 | 69.77 | 64.30 | 69.44 | 55.39 | 54 | 71.54 | 38.35 | 51.59 | 58.44 |
| Bahrain | 62.91 | 78.40 | 67.20 | 68.30 | 86.90 | 68.73 | 65.10 | 66.45 | 71.32 | 46.30 | 64.31 | 38.75 | 34.47 | -0.00 |
| Botswana | 54.23 | 53.69 | 45.50 | 100 | 59 | 56.84 | 52.13 | 60.25 | 59.71 | 39.20 | 53.84 | 31.44 | 40.19 | -46.08 |
| Brazil | 48.06 | 65.45 | 58.10 | 69.40 | 79.40 | 56.42 | 45.88 | 53.46 | 64.63 | 81.30 | 60.23 | 48.90 | 52.98 | -9.20 |
| Bulgaria | 56.80 | 71.34 | 73.40 | 90 | 77.70 | 67.96 | 55.67 | 64.58 | 59.57 | 54.90 | 61.86 | 44.94 | 46.37 | 58.64 |
| Burkina Faso | 48.53 | 34.82 | 26.80 | 75 | 42 | 31.55 | 50.32 | 52.35 | 46.20 | 38.90 | 49.88 | 24.81 | 34.53 | -8.67 |
| Burundi | 40.73 | 39.16 | 14.80 | 61.85 | 43.10 | 36.57 | 47.88 | 50.71 | 47.53 | 22.50 | 53.71 | 24.43 | 33.51 | -39.26 |
| Chile | 63.90 | 76.28 | 63.10 | 100 | 89.70 | 69.82 | 67.99 | 62.79 | 81.98 | 63.20 | 65.26 | 42.50 | 47.63 | 61.59 |
| China | 56.78 | 77.91 | 78.50 | 98.80 | 87.80 | 64.10 | 57.54 | 59.24 | 74.97 | 100 | 66.40 | 64.83 | 72.94 | 65.59 |
| Columbia | 49.25 | 64.31 | 49.90 | 90 | 95 | 60.48 | 52.70 | 59.15 | 64.63 | 66.70 | 64.20 | 36.45 | 55.43 | -17.06 |
| Czech Republic | 60.89 | 83.81 | 68.40 | 100 | 85.60 | 72.88 | 57.35 | 63.29 | 67.58 | 64.80 | 68.68 | 56.90 | 51.35 | 65.88 |
| Denmark | 77.39 | 87.11 | 83.30 | 100 | 92.60 | 85.70 | 66.91 | 78.24 | 86.80 | 59.90 | 79.99 | 76.21 | 39.79 | 65.12 |
| Dominican Republic | 50.06 | 64.91 | 51.80 | 74.95 | 75.70 | 58.69 | 53.70 | 62.89 | 61.56 | 53.80 | 57.10 | 34.62 | 44.18 | -16.16 |
| Ecuador | 47.77 | 69.13 | 47.60 | 73.70 | 85.0 | 61.40 | 43.32 | 51.86 | 56.34 | 54 | 45.74 | 33.01 | 33.47 | -22.76 |
| Egypt | 51.32 | 73.05 | 40.60 | 44.70 | 65.0 | 54.21 | 50.73 | 49.51 | 56.11 | 73.60 | 56.10 | 39.61 | 46.60 | -27.15 |
| El Salvador | 39.85 | 61.02 | 40.60 | 69.75 | 78.10 | 48.43 | 53.93 | 53.40 | 62.24 | 42.90 | 52.70 | 27.91 | 36.08 | -21.09 |
| Estonia | 70.23 | 75.77 | 78.80 | 100 | 84.50 | 79.37 | 61.97 | 70.23 | 65.21 | 42.80 | 69.93 | 52.10 | 40.23 | 71.87 |
| France | 70.04 | 89.73 | 73.70 | 99.85 | 99.20 | 71.94 | 62.23 | 62.93 | 85.87 | 81.60 | 71.39 | 77.18 | 47.96 | 56.85 |
| Gambia, The | 48.53 | 47.37 | 31.40 | 65.45 | 52.30 | 45.03 | 54.24 | 55.04 | 49.66 | 20.60 | 51.05 | 30.49 | 10.07 | -18.08 |
| Georgia | 60.99 | 67.60 | 63.70 | 74.40 | 74.40 | 69.83 | 58.40 | 65.33 | 56.18 | 41.60 | 62.20 | 32.69 | 43.92 | 76.13 |
| Germany | 72.38 | 90.22 | 70.0 | 100 | 92.30 | 84.19 | 68.21 | 72.76 | 79.10 | 86 | 79.54 | 86.82 | 54.17 | 63.05 |
| Ghana | 54.39 | 46.64 | 49.10 | 59.55 | 53.30 | 52.17 | 53.22 | 56.03 | 48.82 | 54.20 | 54.14 | 32.86 | 41.88 | -6.20 |
| Greece | 50.50 | 77.66 | 64.70 | 75.0 | 93.50 | 70.49 | 53.83 | 52.74 | 48.98 | 59.60 | 58.77 | 45.14 | 27.64 | -48.61 |
| Guatemala | 42.45 | 55.86 | 37.70 | 74.85 | 74.0 | 51.39 | 59.00 | 50.92 | 57.52 | 51.20 | 55.80 | 31.56 | 37.84 | -11.26 |
| Guinea | 46.07 | 41.68 | 28.70 | 65.85 | 39.90 | 36.96 | 54.64 | 56.97 | 53.50 | 36.30 | 58.16 | 34.93 | 31.17 | -19.13 |
| Iceland | 74.11 | 76.36 | 85.30 | 100 | 97.70 | 83.42 | 59.03 | 74.93 | 71.32 | 32.30 | 77.05 | 65.12 | 60.82 | 79.86 |
| Indonesia | 58.10 | 67.74 | 55.40 | 90.0 | 70.80 | 64.02 | 58.24 | 57.68 | 63.94 | 82.40 | 69.60 | 37.70 | 59.43 | -7.21 |
| Israel | 65.64 | 83.04 | 67.60 | 100 | 98.10 | 79.61 | 61.79 | 71.09 | 80.56 | 59.80 | 79.55 | 74.17 | 36.77 | -32.29 |
| Italy | 58.56 | 84.09 | 64.50 | 84.65 | 99.60 | 70.40 | 61.88 | 56.58 | 67.58 | 79.30 | 65.74 | 65.53 | 44.66 | 53.30 |
| Jordan | 59.82 | 67.45 | 51.0 | 69.85 | 86.70 | 67.16 | 55.84 | 57.73 | 71.61 | 48.80 | 56.58 | 38.79 | 30.82 | -33.15 |
| Kazakhstan | 55.62 | 68.34 | 68.0 | 86.20 | 71.0 | 67.48 | 55.70 | 67.81 | 53.08 | 63.40 | 66.65 | 32.01 | 61.32 | 65.63 |
| Kenya | 54.66 | 53.61 | 35.70 | 71.75 | 55.10 | 56.30 | 52.87 | 58.86 | 58.04 | 52.70 | 63.95 | 36.30 | 29.33 | -27.42 |
| Latvia | 59.29 | 76.02 | 79.70 | 100 | 76.80 | 76.24 | 58.61 | 67.28 | 57.10 | 44.40 | 65.90 | 42.43 | 31.23 | 65.40 |
| Lebanon | 44.40 | 61.28 | 46.70 | 66.55 | 82.0 | 64.24 | 51.20 | 54.41 | 64.72 | 48.60 | 52.99 | 38.48 | 24.36 | -50.70 |
| Lesotho | 43.00 | 33.26 | 43.0 | 73.80 | 21.70 | 48.31 | 50.30 | 61.28 | 43.38 | 24.80 | 50.13 | 21.82 | 34.86 | 56.74 |
| Madagascar | 39.95 | 31.42 | 21.50 | 69.40 | 48.30 | 38.55 | 47.93 | 53.93 | 46.73 | 40.10 | 51.34 | 25.29 | 36.86 | -26.22 |
| Malawi | 45.73 | 35.54 | 25.20 | 66.15 | 47.0 | 38.18 | 47.87 | 60.08 | 48.78 | 34.20 | 48.76 | 26.85 | 49.66 | -22.75 |
| Malaysia | 68.58 | 78.03 | 71.60 | 100 | 81.20 | 72.54 | 64.75 | 70.16 | 85.31 | 73.40 | 74.63 | 55.01 | 47.49 | -11.03 |
| Mali | 41.49 | 43.88 | 27.90 | 74.90 | 41.0 | 32.79 | 48.07 | 46.08 | 46.33 | 39.90 | 51.78 | 29.01 | 41.20 | -14.84 |
| Malta | 61.34 | 75.02 | 75.50 | 100 | 93.20 | 72.17 | 59.56 | 66.63 | 72.13 | 37.20 | 59.40 | 50.51 | 27.45 | -41.36 |
| Mauritius | 64.69 | 68.70 | 68.30 | 89.45 | 77.40 | 60.63 | 64.46 | 59.04 | 77.19 | 37.20 | 66.15 | 38.06 | 42.41 | -25.78 |
| Mexico | 48.29 | 72.45 | 55.0 | 97.80 | 82.0 | 58.23 | 57.69 | 55.84 | 61.78 | 80.80 | 65.83 | 43.58 | 53.89 | -28.37 |
| Moldova | 51.36 | 66.18 | 66.80 | 73.40 | 71.90 | 61.47 | 54.99 | 61.92 | 46.84 | 36.10 | 60.15 | 29.91 | 44.50 | 61.63 |
| Mongolia | 49.77 | 56.55 | 46.50 | 66.70 | 63.30 | 56.55 | 50.04 | 64.0 | 50.51 | 41.80 | 53.29 | 32.34 | 37.57 | 84.55 |
| Morocco | 60.01 | 72.63 | 46.20 | 90.0 | 72.30 | 48.62 | 55.10 | 51.50 | 67.47 | 60.50 | 59.80 | 35.12 | 25.66 | -36.66 |
| Mozambique | 39.32 | 35.15 | 23.10 | 42.35 | 33.10 | 30.26 | 46.74 | 43.14 | 48.46 | 41.10 | 46.76 | 27.44 | 29.89 | -30.10 |
| Namibia | 56.81 | 58.50 | 48.10 | 72.15 | 53.40 | 54.60 | 53.56 | 63.71 | 69.09 | 36.70 | 51.21 | 35.64 | 37.40 | -36.67 |
| Nepal | 47.89 | 51.79 | 38.60 | 73.95 | 65.90 | 49.32 | 43.0 | 49.16 | 66.37 | 47.70 | 55.75 | 29.41 | 41.95 | -48.47 |
| Netherlands | 78.57 | 94.34 | 76.30 | 100 | 94.20 | 84.63 | 69.93 | 74.91 | 84.62 | 74.30 | 80.59 | 76.31 | 41.37 | 62.20 |
| Norway | 76.92 | 75.81 | 83.10 | 100 | 94.50 | 83.77 | 60.86 | 73.32 | 82.04 | 61.40 | 76.90 | 68.01 | 50.80 | 72.40 |
| Oman | 62.34 | 80.51 | 58.10 | 67.40 | 80.70 | 71.54 | 63.13 | 55.77 | 63.90 | 55.90 | 62.83 | 41.25 | 42.31 | -22.21 |
| Pakistan | 47.70 | 55.56 | 25.20 | 68.75 | 56.30 | 40.73 | 45.50 | 51.27 | 55.04 | 71.20 | 63.31 | 35.76 | 46.32 | -32.34 |
| Paraguay | 44.28 | 59.83 | 45.70 | 74.80 | 81.40 | 50.79 | 54.61 | 55.20 | 56.04 | 47.30 | 51.24 | 22.42 | 31.81 | -19.16 |
| Peru | 48.86 | 62.30 | 45.70 | 100 | 94.60 | 60.21 | 57.08 | 59.02 | 61.44 | 62.20 | 55.81 | 32.74 | 48.63 | -30.80 |
| Philippines | 49.98 | 57.83 | 49.70 | 89.95 | 65.60 | 63.73 | 57.75 | 64.94 | 68.32 | 71.0 | 65.73 | 37.96 | 50.33 | -12.20 |
| Poland | 56.42 | 81.14 | 65.40 | 100 | 83.80 | 72.13 | 58.13 | 59.89 | 64.06 | 74.10 | 62.01 | 49.66 | 49.92 | 65.31 |
| Portugal | 64.52 | 83.60 | 71.20 | 85.0 | 94.20 | 70.01 | 59.76 | 63.18 | 70.04 | 60.50 | 69.70 | 53.69 | 42.01 | -40.87 |
| Romania | 58.07 | 71.68 | 72.0 | 89.65 | 77.20 | 62.48 | 55.39 | 61.57 | 56.98 | 65.20 | 59.68 | 42.33 | 52.44 | 64.61 |
| Russian Federation | 52.57 | 73.85 | 77.0 | 90.0 | 69.20 | 68.30 | 52.91 | 61.03 | 55.66 | 84.20 | 63.11 | 52.93 | 61.85 | 100 |
| Rwanda | 63.22 | 52.0 | 37.60 | 72.65 | 61.40 | 40.13 | 55.35 | 63.58 | 56.34 | 35.10 | 65.59 | 30.93 | 39.07 | -46.58 |
| Saudi Arabia | 63.20 | 78.05 | 69.30 | 100 | 82.20 | 75.33 | 64.92 | 56.63 | 70.69 | 76.30 | 53.11 | 50.56 | 62.00 | -25.95 |
| Spain | 65.07 | 90.31 | 78.20 | 90.0 | 100 | 71.57 | 61.01 | 61.10 | 77.51 | 77.0 | 67.31 | 64.33 | 48.97 | -48.94 |
| Sri Lanka | 51.60 | 69.23 | 40.30 | 68.0 | 87.10 | 63.77 | 43.26 | 51.76 | 56.97 | 58.40 | 60.04 | 34.90 | 49.29 | -8.95 |
| Sweden | 75.21 | 84.01 | 87.80 | 100 | 96.60 | 83.72 | 66.29 | 69.38 | 88.03 | 65.40 | 79.44 | 79.09 | 48.77 | 82.54 |
| Switzerland | 77.52 | 93.15 | 78.60 | 100 | 99.90 | 86.73 | 63.80 | 79.48 | 89.72 | 66.20 | 71.56 | 81.20 | 58.01 | 71.37 |
| Thailand | 54.83 | 67.84 | 60.10 | 90.0 | 88.90 | 62.33 | 53.48 | 63.39 | 85.07 | 75.50 | 71.96 | 43.87 | 57.88 | -11.19 |
| Turkey | 53.92 | 74.29 | 57.80 | 61.30 | 87.10 | 60.83 | 54.10 | 52.88 | 61.20 | 79.0 | 58.81 | 44.50 | 56.95 | 55.73 |
| Uganda | 48.03 | 47.88 | 29.40 | 74.15 | 53.0 | 42.26 | 49.07 | 59.96 | 50.30 | 47.40 | 56.35 | 29.54 | 32.34 | -33.31 |
| Ukraine | 47.85 | 70.34 | 51.90 | 57.90 | 65.60 | 69.91 | 56.51 | 61.38 | 42.30 | 63.0 | 57.15 | 40.12 | 40.81 | 58.16 |
| United Kingdom | 74.42 | 88.88 | 73.0 | 100 | 91.60 | 81.92 | 64.60 | 74.97 | 88.13 | 81.80 | 77.01 | 78.16 | 49.03 | 64.81 |
| United States | 71.17 | 87.91 | 74.30 | 99.75 | 83.0 | 82.48 | 68.55 | 77.98 | 90.98 | 99.50 | 84.23 | 84.14 | 60.43 | 65.05 |
| Zambia | 45.17 | 43.27 | 34.20 | 64.05 | 47.30 | 47.65 | 48.58 | 49.73 | 47.83 | 45.40 | 56.46 | 28.56 | 52.26 | -24.86 |

Note: the pillar values are rounded off to two decimal places in the table, while exact measurements are used in calculation of GACI.

**Appendix Table 3: Validity of Global Agricultural Competitiveness Index**

The final step of an index construction is the index validation. Validation is done in order to confirm the correctness of the measure used. There are different methods used for validation, among which item analysis and predicting the related measures are the commonly used methods. During the item analysis, the level of the relationship between the composite index and the individual items included in it, is considered. While in the later, the accuracy of the composite index in predicting the related measures is checked. Item analysis is used for checking the validity of GACI. An analysis of bivariate correlation, also known as Pearson correlation, can be used to address test validity. To determine the validity of the measurements, Pearson Correlation is used.

It can be interpreted from the table that all the pillars/items in the index measure the same phenomenon as overall index does. Hence pillars which are utilized in the index are qualified to be a part of the final index. The correlation between each indicator and the GACI is explained in Table, which identifies a significant result at either the 0.01 level or the 0.05 level, with only one observation significant the 0.10 level. Moreover, the correlation of each pillar with GACI is also sufficiently high, ranging between 0.58 and 0.91, to validate this index. Thus, each pillar and GACI in itself are proved valid.

## Table 3: Validity Test using Pearson correlation

| Variables | (1) | (2) | (3) | (4) | (5) | (6) | (7) | (8) | (9) | (10) | (11) | (12) | (13) | (14) | (15) |
|---|---|---|---|---|---|---|---|---|---|---|---|---|---|---|---|
| Pillar 1 | 1.000 | | | | | | | | | | | | | | |
| Pillar 2 | 0.797* | 1.000 | | | | | | | | | | | | | |
| | (0.000) | | | | | | | | | | | | | | |
| Pillar 3 | 0.808* | 0.864* | 1.000 | | | | | | | | | | | | |
| | (0.000) | (0.000) | | | | | | | | | | | | | |
| Pillar 4 | 0.693* | 0.619* | 0.687* | 1.000 | | | | | | | | | | | |
| | (0.000) | (0.000) | (0.000) | | | | | | | | | | | | |
| Pillar 5 | 0.666* | 0.878* | 0.779* | 0.586* | 1.000 | | | | | | | | | | |
| | (0.000) | (0.000) | (0.000) | (0.000) | | | | | | | | | | | |
| Pillar 6 | 0.822* | 0.895* | 0.904* | 0.639* | 0.828* | 1.000 | | | | | | | | | |
| | (0.000) | (0.000) | (0.000) | (0.000) | (0.000) | | | | | | | | | | |
| Pillar 7 | 0.823* | 0.739* | 0.722* | 0.689* | 0.606* | 0.752* | 1.000 | | | | | | | | |
| | (0.000) | (0.000) | (0.000) | (0.000) | (0.000) | (0.000) | | | | | | | | | |
| Pillar 8 | 0.811* | 0.613* | 0.727* | 0.665* | 0.509* | 0.765* | 0.739* | 1.000 | | | | | | | |
| | (0.000) | (0.000) | (0.000) | (0.000) | (0.000) | (0.000) | (0.000) | | | | | | | | |
| Pillar 9 | 0.822* | 0.753* | 0.695* | 0.715* | 0.724* | 0.717* | 0.744* | 0.654* | 1.000 | | | | | | |
| | (0.000) | (0.000) | (0.000) | (0.000) | (0.000) | (0.000) | (0.000) | (0.000) | | | | | | | |
| Pillar 10 | 0.375* | 0.640* | 0.481* | 0.385* | 0.535* | 0.483* | 0.356* | 0.201 | 0.529* | 1.000 | | | | | |
| | (0.001) | (0.000) | (0.000) | (0.001) | (0.000) | (0.000) | (0.001) | (0.078) | (0.000) | | | | | | |
| Pillar 11 | 0.842* | 0.754* | 0.736* | 0.685* | 0.645* | 0.779* | 0.771* | 0.787* | 0.783* | 0.512* | 1.000 | | | | |
| | (0.000) | (0.000) | (0.000) | (0.000) | (0.000) | (0.000) | (0.000) | (0.000) | (0.000) | (0.000) | | | | | |
| Pillar 12 | 0.844* | 0.826* | 0.770* | 0.661* | 0.702* | 0.803* | 0.723* | 0.694* | 0.826* | 0.622* | 0.830* | 1.000 | | | |
| | (0.000) | (0.000) | (0.000) | (0.000) | (0.000) | (0.000) | (0.000) | (0.000) | (0.000) | (0.000) | (0.000) | | | | |
| Pillar 13 | 0.335* | 0.413* | 0.448* | 0.430* | 0.338* | 0.379* | 0.270* | 0.311* | 0.367* | 0.652* | 0.456* | 0.426* | 1.000 | | |
| | (0.003) | (0.000) | (0.000) | (0.000) | (0.002) | (0.001) | (0.017) | (0.006) | (0.001) | (0.000) | (0.000) | (0.000) | | | |
| Pillar 14 | 0.474* | 0.458* | 0.585* | 0.407* | 0.265* | 0.520* | 0.430* | 0.576* | 0.254* | 0.264* | 0.463* | 0.489* | 0.468* | 1.000 | |
| | (0.000) | (0.000) | (0.000) | (0.000) | (0.019) | (0.000) | (0.000) | (0.000) | (0.025) | (0.020) | (0.000) | (0.000) | (0.000) | | |
| GACI | 0.863* | 0.898* | 0.912* | 0.764* | 0.778* | 0.903* | 0.790* | 0.786* | 0.792* | 0.624* | 0.855* | 0.891* | 0.581* | 0.698* | 1.000 |
| | (0.000) | (0.000) | (0.000) | (0.000) | (0.000) | (0.000) | (0.000) | (0.000) | (0.000) | (0.000) | (0.000) | (0.000) | (0.000) | (0.000) | |

** Shows significance at p<.05*

## Appendix: Table 4: Global Competitiveness Index Scores and Rankings

The Global Competitiveness Index (GCI) is an already developed index with its scores and rankings already calculated. However, for the purpose of the present study, those countries were picked for which the GACI scores are calculated. The selected countries are then re ranked within the 78 countries on the basis of their GCI scores and then compared with the GACI scores.

**Table 4: GCI scores and ranking within the 78 selected countries.**

| Country | GCI Ranking | GCI Scores2019 |
|---|---|---|
| United States | 1 | 83.700 |
| Netherlands | 2 | 82.400 |
| Switzerland | 3 | 82.300 |
| Germany | 4 | 81.800 |
| Denmark | 5 | 81.200 |
| Sweden | 6 | 81.200 |
| United Kingdom | 7 | 81.200 |
| France | 8 | 78.800 |
| Australia | 9 | 78.700 |
| Norway | 10 | 78.100 |
| Israel | 11 | 76.700 |
| Austria | 12 | 76.600 |
| Spain | 13 | 75.300 |
| Iceland | 14 | 74.700 |
| Malaysia | 15 | 74.600 |
| China | 16 | 73.900 |
| Italy | 17 | 71.500 |
| Czech Republic | 18 | 70.900 |
| Estonia | 19 | 70.900 |
| Chile | 20 | 70.500 |
| Portugal | 21 | 70.400 |
| Saudi Arabia | 22 | 70 |
| Poland | 23 | 68.900 |
| Malta | 24 | 68.500 |
| Thailand | 25 | 68.100 |
| Latvia | 26 | 67 |
| Russian Federation | 27 | 66.700 |
| Bahrain | 28 | 65.400 |
| Bulgaria | 29 | 64.900 |
| Mexico | 30 | 64.900 |
| Indonesia | 31 | 64.600 |
| Romania | 32 | 64.400 |
| Mauritius | 33 | 64.300 |
| Oman | 34 | 63.600 |
| Kazakhstan | 35 | 62.900 |
| Azerbaijan | 36 | 62.700 |
| Columbia | 37 | 62.700 |
| Greece | 38 | 62.600 |
| Turkey | 39 | 62.100 |
| Philippines | 40 | 61.900 |
| Peru | 41 | 61.700 |
| Armenia | 42 | 61.300 |
| Brazil | 43 | 60.900 |
| Jordan | 44 | 60.900 |
| Georgia | 45 | 60.600 |
| Morocco | 46 | 60 |
| Dominican Republic | 47 | 58.300 |
| Albania | 48 | 57.600 |
| Argentina | 49 | 57.200 |
| Sri Lanka | 50 | 57.100 |
| Ukraine | 51 | 57 |
| Moldova | 52 | 56.700 |
| Lebanon | 53 | 56.300 |
| Ecuador | 54 | 55.700 |
| Botswana | 55 | 55.500 |
| Egypt | 56 | 54.500 |
| Namibia | 57 | 54.500 |
| Kenya | 58 | 54.100 |
| Paraguay | 59 | 53.600 |
| Guatemala | 60 | 53.500 |
| Rwanda | 61 | 52.800 |
| El Salvador | 62 | 52.600 |
| Mongolia | 63 | 52.600 |
| Nepal | 64 | 51.600 |
| Pakistan | 65 | 51.400 |
| Ghana | 66 | 51.200 |
| Uganda | 67 | 48.900 |
| Zambia | 68 | 46.500 |
| Guinea | 69 | 46.100 |
| Gambia, The | 70 | 45.900 |
| Malawi | 71 | 43.700 |
| Mali | 72 | 43.600 |
| Burkina Faso | 73 | 43.400 |
| Lesotho | 74 | 42.900 |
| Madagascar | 75 | 42.900 |
| Burundi | 76 | 40.300 |
| Angola | 77 | 38.100 |
| Mozambique | 78 | 38.100 |

**Appendix Table 5: Comparison between GACI and GCI Scores**

A comparison and difference between of the GACI and GCI scores is shown in the following Table.

| Country | GACI Scores 2019 | GCI Scores 2019 | GACI-GCI |
|---|---|---|---|
| Albania | 47.963 | 57.600 | -9.637 |
| Angola | 33.496 | 38.100 | -4.604 |
| Argentina | 48.087 | 57.200 | -9.113 |
| Armenia | 60.191 | 61.300 | -1.109 |
| Australia | 68.512 | 78.700 | -10.188 |
| Austria | 74.268 | 76.600 | -2.332 |
| Azerbaijan | 61.622 | 62.700 | -1.078 |
| Bahrain | 58.509 | 65.400 | -6.891 |
| Botswana | 47.138 | 55.500 | -8.362 |
| Brazil | 55.358 | 60.900 | -5.542 |
| Bulgaria | 63.123 | 64.900 | -1.777 |
| Burkina Faso | 39.073 | 43.400 | -4.327 |
| Burundi | 34.088 | 40.300 | -6.212 |
| Chile | 68.267 | 70.500 | -2.233 |
| China | 73.242 | 73.900 | -0.658 |
| Columbia | 56.510 | 62.700 | -6.190 |
| Czech Republic | 69.100 | 70.900 | -1.800 |
| Denmark | 77.075 | 81.200 | -4.125 |
| Dominican Republic | 51.986 | 58.300 | -6.314 |
| Ecuador | 48.541 | 55.700 | -7.159 |
| Egypt | 48.142 | 54.500 | -6.358 |
| El Salvador | 46.131 | 52.600 | -6.469 |
| Estonia | 68.785 | 70.900 | -2.115 |
| France | 75.033 | 78.800 | -3.767 |
| Gambia, The | 38.797 | 45.900 | -7.103 |
| Georgia | 60.526 | 60.600 | -0.074 |
| Germany | 78.481 | 81.800 | -3.319 |
| Ghana | 46.435 | 51.200 | -4.765 |
| Greece | 52.139 | 62.600 | -10.461 |
| Guatemala | 47.773 | 53.500 | -5.727 |
| Guinea | 40.408 | 46.100 | -5.692 |
| Iceland | 74.094 | 74.700 | -0.606 |
| Indonesia | 59.131 | 64.600 | -5.469 |
| Israel | 66.102 | 76.700 | -10.598 |
| Italy | 68.312 | 71.500 | -3.188 |
| Jordan | 52.071 | 60.900 | -8.829 |
| Kazakhstan | 63.017 | 62.900 | 0.117 |
| Kenya | 46.554 | 54.100 | -7.546 |
| Latvia | 64.313 | 67 | -2.687 |
| Lebanon | 46.373 | 56.300 | -9.927 |
| Lesotho | 43.313 | 42.900 | 0.413 |
| Madagascar | 37.506 | 42.900 | -5.394 |
| Malawi | 39.375 | 43.700 | -4.325 |
| Malaysia | 66.547 | 74.600 | -8.053 |
| Mali | 39.248 | 43.600 | -4.352 |
| Malta | 57.768 | 68.500 | -10.732 |
| Mauritius | 56.279 | 64.300 | -8.021 |
| Mexico | 57.486 | 64.900 | -7.414 |
| Moldova | 56.225 | 56.700 | -0.475 |
| Mongolia | 53.819 | 52.600 | 1.219 |
| Morocco | 50.653 | 60 | -9.347 |
| Mozambique | 32.623 | 38.100 | -5.477 |
| Namibia | 46.727 | 54.500 | -7.773 |
| Nepal | 43.736 | 51.600 | -7.864 |
| Netherlands | 78.018 | 82.400 | -4.382 |
| Norway | 75.702 | 78.100 | -2.398 |
| Oman | 55.961 | 63.600 | -7.639 |
| Pakistan | 45.022 | 51.400 | -6.378 |
| Paraguay | 46.875 | 53.600 | -6.725 |
| Peru | 54.128 | 61.700 | -7.572 |
| Philippines | 55.758 | 61.900 | -6.142 |
| Poland | 67.284 | 68.900 | -1.616 |
| Portugal | 60.467 | 70.400 | -9.933 |
| Romania | 63.520 | 64.400 | -0.880 |
| Russian Federation | 68.757 | 66.700 | 2.057 |
| Rwanda | 44.741 | 52.800 | -8.059 |
| Saudi Arabia | 62.596 | 70 | -7.404 |
| Spain | 64.532 | 75.300 | -10.768 |
| Sri Lanka | 51.832 | 57.100 | -5.268 |
| Sweden | 79.019 | 81.200 | -2.181 |
| Switzerland | 79.803 | 82.300 | -2.497 |
| Thailand | 61.712 | 68.100 | -6.388 |
| Turkey | 61.314 | 62.100 | -0.786 |
| Uganda | 41.883 | 48.900 | -7.017 |
| Ukraine | 55.923 | 57 | -1.077 |
| United Kingdom | 77.738 | 81.200 | -3.462 |
| United States | 80.676 | 83.700 | -3.024 |
| Zambia | 41.828 | 46.500 | -4.672 |